\documentclass[letter]{aa} 

\usepackage{graphicx}
\usepackage{amsmath}	
\usepackage{amssymb}	
\usepackage{multirow}
\usepackage{booktabs}
\usepackage{textcomp,gensymb}
\usepackage{xcolor,color}
\usepackage[caption = false]{subfig}
\usepackage{ulem}
\usepackage[cm]{aeguill}
\usepackage{graphicx}
\usepackage[]{url}
\usepackage[T1]{fontenc}
\usepackage{ae,aecompl}
\usepackage{tabularx}
\usepackage{supertabular}
\usepackage{longtable}
\usepackage{ntheorem}
\usepackage{hyperref}

\makeatletter
\renewcommand*\aa@pageof{, page \thepage{} of \pageref*{LastPage}}
\makeatother

\usepackage{txfonts}
\begin{document}

   \title{Ready for O4 II: GRANDMA Observations of \textit{Swift} GRBs during eight-weeks of Spring 2022}
   \titlerunning{Ready for O4 II}

   \author{I.~Tosta~e~Melo$^{1}$\thanks{Corresponding authors; E-mail: iara.tosta.melo@dfa.unict.it, ducoin@iap.fr}, J.-G.~Ducoin$^{2\star}$, Z.~Vidadi$^{3}$, C.~Andrade$^{4}$, V.~Rupchandani$^{5}$, S.~Agayeva$^{3}$, J.~Abdelhadi$^{6}$, L. ~Abe$^{7}$, O.~Aguerre-Chariol$^{8}$, V.~Aivazyan$^{9,10}$
% ...
}
   \authorrunning{GRANDMA consortium}

   \institute{}

   \date{Received XXX; accepted YYY}

 
  \abstract
   {}
   {\textcolor{black}{We present a campaign designed to train the GRANDMA network and its infrastructure to follow up on transient alerts and detect their early afterglows.
   In preparation for O4 II campaign, we focused on GRB alerts as they are expected to be an electromagnetic counterpart of gravitational-wave events. 
   Our goal was to improve our response to the alerts and start prompt observations as soon as possible to better prepare the GRANDMA network for the fourth observational run of LIGO-Virgo-Kagra (which started at the end of May 2023), and future missions such as SM.}}
   {\textcolor{black}{To receive, manage and send out observational plans to our partner telescopes we set up dedicated infrastructure and a rota of follow-up adcates were organized to guarantee round-the-clock assistance to our telescope teams. 
   To ensure a great number of observations, we focused on \textit{Swift} GRBs whose localization errors were generally smaller than the GRANDMA telescopes' field of view. This allowed us to bypass the transient identification process and focus on the reaction time and efficiency of the network.}}
   {\textcolor{black}{During 'Ready for O4 II', 11 \textit{Swift}/INTEGRAL GRB triggers were selected, nine fields had been observed, and three afterglows were detected (GRB\,220403B, GRB\,220427A, GRB\,220514A), with 17 GRANDMA telescopes and 17 amateur astronomers from the citizen science project Kilonova-Catcher.  
    Here we highlight the GRB\,220427A analysis where our long-term follow-up of the host galaxy allowed us to obtain a photometric redshift of $z=0.82\pm0.09$, its lightcurve elution, fit the decay slope of the afterglows, and study the properties of the host galaxy.}}
  {\textcolor{black}{During this 8-week-long GRB follow-up campaign, we successfully fulfilled our goal of training telescope teams for O4 and the improvement of the associated technical toolkits. 
  For seven of the GRB alerts, our network was able to start the first observations less than one hour after the GRB trigger time. We also characterized the network efficiency to observe GRB afterglow given the resulting time delay and limiting magnitude as well as to study its lightcurve elution through the observation of GRB\, 220427A.}}


   \maketitle
%

\section{Introduction}

The detection and successful electromagnetic characterization of GW170817 by the International Gravitational-Wave Observatory Network (IGWN)  \citep{LSC_BNS_2017PhRvL, LSC_MM_2017ApJ} boosted the importance of dedicated multi-messenger efforts in time-domain astronomy. 
In particular, the identification and follow-up studies of GRB170817A by Fermi-GBM \cite{goldstein_ordinary_2017} and INTEGRAL \cite{savchenko_integral_2017} have corroborated the long-predicted connection between binary neutron stars (hereafter, BNS) and short gamma-ray bursts (sGRBs) (see e.g. \citet{2017ApJ...848L..21A, 2017ApJ...848L..25H, Hallinan_2017Sci} amongst others).
Our current efforts are focused on characterizing the electromagnetic emissions associated with gravitational waves (GWs) produced by compact binaries across various wavelengths. This is done to comprehensively investigate the complete range of astrophysical information related to a compact binary coalescence.The kilonova emission from such sources (eg. \citet{Andreoni_2017PASA, 2017Natur.551..210A, 2017SciBu..62.1433H}) is widely studied to constrain the neutron star equation of state (EOS) \citep{2017ApJ...850L..34B, 2017ApJ...850L..19M, CoDi2018, CoDi2018b, Coughlin2019}, the Hubble constant (eg. \citet{LSC_BNS_2017PhRvL, 2019NatAs...3..940H, PhysRevResearch.2.022006, 2020NatCo..11.4129C}) and studies of $r$-process nucleosynthesis \citep{2021hgwa.bookE..13P}.
In this context, the GRANDMA (Global Rapid Advanced Network Deted to the Multi-messenger Addicts; \citealt{GRANDMAO3A,GRANDMA03B}) collaboration's primary goal is to follow-up GW alerts provided by the IGWN with telescopes distributed worldwide in order to achieve as short a response time as possible. 
GRANDMA brings together telescopes located in both hemispheres with the objective of coordinating them all as a single facility to respond to GW (or other transient) alerts. 
Coordinated observations can be achieved thanks to a common scheduler that automatically sends coordinated observation plans to individual telescopes. 
A centralized data cloud coupled with a database allows the storage of images that are uploaded by telescope teams within GRANDMA's network. The images provided by the team will then be processed by tools used within the collaboration, see \ref{datareduc}.
GRANDMA performed an intensive follow-up of GW alerts during the O3 LIGO-Virgo run and demonstrated an efficient response time: less than 90 minutes of latency for 50\% of the alerts, with a record of 15 minutes (\citealt{GRANDMAO3A,GRANDMA03B}). 
After the end of O3 in March 2020, GRANDMA has continued on its path, increasing in size through the introduction of a number of new groups.
In order to facilitate both software and educational advancement, multiple observation campaigns were conducted, typically on an annual basis. These campaigns served a dual purpose. They prepared and trained the collaboration, which included both telescope teams and infrastructure development, while also allowing for the creation of novel tools within the context of \texttt{SkyPortal}\footnote{https://skyportal.io/)} \citep{CoBl2023}. The ultimate aim was to be well-prepared for O4, which commenced at the start of May 2023 with a planned duration of 18 months.
The first GRANDMA ``training campaign'' occurred from April to September 2021 and was dedicated to the follow-up of transient candidates detected by the Zwicky Transient Facility (ZTF) and processed by the Fink broker (\citealt{GRANDMA_ztf_fink,2021MNRAS.501.3272M,2014htu..conf...27B}). 
This work presented the first integration of amateur astronomers, through the citizen science program ``Kilonova Catcher'' initiated by GRANDMA.
Lastly, follow-up observations dedicated to GRB221009A were performed with data from the prompt emission 30 days after the GRB alerts \citep{2023arXiv230206225K}. 
In this paper, we present the results of a ``training GRANDMA campaign'' with a duration of eight-weeks. 
We followed the GRB observed by \textit{Swift} due to its small localization error which is consistent with GRANDMA telescopes' field of view.
Besides preparing our network for O4, we tested our capability to identify and characterize GRB afterglows, in the framework of the forth-coming mission SM \citep{2016arXiv161006892W}.
\textcolor{black}{GRANDMA was able to observe 9 of the 11 GRBs detected by \textit{Swift}/INTEGRAL during the campaign period, corresponding to a success of $\sim$\,82\% with three afterglow detections.}
We present, in more detail, the observation and detection of an afterglow, GRB\,220427A, which was observed at very early times and followed up until it was host-galaxy dominated. 
The current study demonstrates the effectiveness of GRANDMA in swiftly initiating observations upon the detection of a transient alert, while also promptly observing their optical counterparts. \textcolor{black}{This is crucial to ensuring that no gravitational wave (GW) event with expected electromagnetic (EM) observations is missed.}
The article is structured as follows: Section ~\ref{grandmaknc} details the existing GRANDMA network, Section ~\ref{finkKn} outlines the infrastructure established for the GRB campaign, Sections ~\ref{readyforO4} and ~\ref{results} present the outcomes of our campaign, and Section ~\ref{conclusions} summarizes our conclusions.

\section{Motivations}

For its second campaign between the 3rd and 4th observational runs of the IGWN, the GRANDMA collaboration performed a GRB follow-up campaign lasting eight weeks, from March 19th to {\textcolor{black}{May 14th}, 2022. 
The 'Ready for O4 II' campaign was designed to measure and improve the reaction of the teams belonging to the network as well as their performance in following up fast transients. Special attention was given to the new teams (such as Pico Dos Dias and SOAR Observatories) in the collaboration. The list and description of the teams who joined the collaboration for this run are presented in Appendix \ref{tel}. 

Given this objective, we chose to focus on \textit{Swift} GRBs. 
This provided us the opportunity to easily quantify the reaction time and efficiency of the network as the \textit{Swift} typical localization errors are generally smaller than the GRANDMA telescopes' field of view. This aided the complication of the transient identification process in large localization errors that have been previously tested and used during the GRANDMA O3 follow-up \citep{GRANDMAO3A, 2020MNRAS.497.5518A}. 
Also, at the end of the campaign, on May 14th 2022, a long GRB was detected by INTEGRAL, which was observed by GRANDMA telescopes.

The campaign had several objectives. Primarily, it aimed to gauge the effectiveness of the network by analyzing the response time to the detection of optical/NIR afterglows. The goal was to trigger observations among the telescope teams in our collaboration as early as possible and with the best time sampling achievable, while also extending the observations for as long as the source remains detectable.This approach allowed for a comprehensive analysis of the geometry and physical characteristics of the observed GRBs. A second key goal was to utilize the network for identifying kilonova or supernova counterparts. Additionally, the campaign sought to identify any long-term emissions, such as {\textcolor{black}{supernovae} or other components distinct from the afterglow. Lastly, it aimed to discover and characterize the host galaxy by determining the spectroscopic redshift, especially in the case of a bright host. This task necessitated revisiting the field, potentially employing various filters, at later time points, typically after $\sim$1 month.

The above scientific objectives have been designed to stress the GRANDMA system in various ways to serve as a pathfinder for the O4 run. Among the technical and practical objectives of this campaign, we can specifically mention the desire to optimize and harmonize image collection, processing and data reduction with pipelines built within the collaboration through {\it{Skyportal}} \citep{CoBl2023}. Finally, we aim to improve the overall expertise in time-domain astronomy and the tools of the collaboration dedicated to enabling rapid decision making which is essential for transient follow-up. 
We also wanted to ensure our capacity of person power and technical investment for GRB programs, as the SM (Space-based multi-band astronomical Variable Objects Monitor) mission does for example.


\section{GRANDMA and Kilonova catcher}
\label{grandmaknc}

The GRANDMA consortium is a world-wide network of 18 observatories and {\textcolor{black}{26 telescopes}, 42 institutions, and groups from 18 countries. These facilities make available large amounts of observing time that can be allocated for photometric and/or spectroscopic follow-up of transients. The network has access to wide field-of-view telescopes (FoV $>1$ deg$^2$) located on three continents, and remote and robotic telescopes with narrower fields-of-view \footnote{https://grandma.ijclab.in2p3.fr/}. 
To complement this telescope network, GRANDMA has initiated a citizen science project called Kilonova-Catcher\footnote{\url{http://kilonovacatcher.in2p3.fr/}} \citealt[(KNC)]{GRANDMA03B,GRANDMA_ztf_fink,2023arXiv230206225K}. 
This program allows any interested amateur astronomer to connect to the GRANDMA alert system to perform follow-up observations of promising multi-messenger transient sources. At present, 110 amateur astronomers all around the world are connected to the KNC web interface to receive alerts and customized observation plans as well as subscribe to our mailing list.\\


\section{Interface and Communication for Addicts of the Rapid follow-up in multi-messenger Era}
\label{finkKn}




\subsection{GRANDMA infrastructure}

For this campaign, we set up dedicated infrastructure to receive, manage and send observational plans to our partner telescopes. 
The full code is accessible in the \href{https://github.com/mcoughlin/gwemopt/tree/master/ToO}{gwemopt library} and inherits the same protocol as described in \citet{GRANDMAO3A}. 
We enabled the reception of \href{https://GCN.gsfc.nasa.gov/GCN3_archive.html}{GCN-notice} alerts from \textit{Swift} (i.e. those named ``\textit{Swift}/BAT alert'' or ``\textit{Swift}/BAT position,'' which correspond to packet\_type 97 and packet\_type 61 respectively), and \textit{Swift}-XRT (``Flight \textit{Swift}/XRT Position,'' packet\_type 67). 
\href{https://www.swift.psu.edu/guano/}{BAT-GUANO} triggers were not included in this campaign. 
We recorded the date and time of the GRB discovery, the coordinates and its error, the sky localization and distance from the Sun and Moon, the probability that the alert was astrophysical, and the signal-to-noise ratio of the source in the 2D gamma-ray image. Using these data, we calculated a score of interest, with points for: 1) if the probability to be astrophysical is 1. ; 2) (1) $+$ the moon illumination is below 0.7, and the Moon and Sun distance from the source is more than 20 degrees ; 3) (1)$+$(2)$+$ $SNR_\textrm{image}$ > 6.5. 
We then created a VO event dedicated to each telescope and broadcast via an IP address. 
The  event (described in the Appendix, \ref{tab:KN_observations}) contains the name of the GRB, trigger ID, the trigger time, event status, internal pack number, the long/short classification, the $h_\textbf{ratio}$ of the GRB (if available), the GRANDMA telescope receiver and the coordinates in J2000 with the error. The astronomical teams are listening constantly to our IP address and receive an initial GRANDMA notice within a few seconds, when the \textit{Swift} coordinates are available, followed (if needed) by a second ``update'' GRANDMA notice when \textit{Swift}-XRT coordinates are available (typically within a few tens of minutes). In case of a GRB follow-up, the same coordinates are transmitted to all astronomical teams, but the code allows us to send multiple targets and a dedicated list per telescope receiver.
In this way, telescope teams belonging to our collaboration are requested to make observations. 

For each GRB, our software also automatically creates a repository on our \href{https://grandma-owncloud.lal.in2p3.fr/}{owncloud platform} where the information about the GRB is stored, such as: folders containing information like the observability map, logbooks for campaign members to follow up on and detail the progress of the GRB observations, folders to upload stacked images and photometric data, etc. 
We have implemented redundancy and automated restart to prevent any hardware interruption. 

{\textcolor{black}{During the entire period of the campaign, we did not report any problem with the hardware and observation plans were automatically sent to the astronomical teams successfully}.
In Table~\ref{tab:GRANDMAdelayinfrast}, we give the latency performance of our infrastructure per \textit{Swift} GRB.


\begin{table}
\caption{The delay of GRANDMA notice delivery compared to the trigger time. The delay is calculated from the telescope receiver and takes into account the delay of transmission from \textit{Swift} to send the position of the GRB, the delay of treatment by the GRANDMA infrastructure, and the delay of transmission.}
\addtolength{\tabcolsep}{-4pt}
\begin{tabular}{cccc}
\hline
\hline

Name & Trigger time & BAT pos. & TCA Receiver  \\
 & (UTC) & delay (s) & delay (s) \\
\hline
GRB220319A & 2022-03-19T17:40:33 & 19 & $<1$ \\ 
GRB220325A & 2022-03-25T17:16:23 & 13 & 9 \\ 
GRB220403B & 2022-04-03T20:42:39 & 18 & 1 \\ 
GRB220404A & 2022-04-04T11:54:30 & 53 & $<1$ \\ 
GRB220408A & 2022-04-08T05:46:04 & 12 & $<1$ \\ 
GRB220412A & 2022-04-12T06:36:50 & 54 & $<1$ \\ 
GRB220412B & 2022-04-12T17:06:48 & 13 & $<1$ \\ 
GRB220427A & 2022-04-27T21:00:34 & 25 & $<1$ \\ 
GRB220430A & 2022-04-30T13:53:15 & 275 & 2 \\ 
GRB220501A & 2022-05-01T-19:51:50 & 21 & 70\\ 
\hline
\end{tabular}
\label{tab:GRANDMAdelayinfrast}
\addtolength{\tabcolsep}{4pt}
\end{table}

\subsection{GRANDMA follow-up organization}

In case of an alert, as with any GRANDMA campaign, a rota of follow-up adcates (FA) was organized to guarantee round-the-clock assistance to our telescope teams. 
FAs were needed to guarantee that everything ran smoothly day after day during a campaign and to notify teams of the GRB alerts \textcolor{black}{received} through GCN via the dedicated \textcolor{black}{communication} channel. 
As soon as a GRB alert was received, FAs would alert the teams within minutes post the alert, validating whether or not a given alert should be followed. It is also the role of the FAs to ensure that the dedicated repo for a given alert was correctly made (OwnCloud, $https://owncloud.com/$). 
As soon as the observation plans for each alert were sent to the observers, the FA filled out a logbook reporting whether a telescope could observe and checked if images were correctly uploaded for further analysis.
The FAs must keep the teams up-to-date about all circulars published about a given alert (reported via GCN), to decide whether or not to stop observations. 
In case of a counterpart discovery by external telescopes and reported via GCN, the FA had the responsibility to ask for a new
observation plan to follow up on this transient.
The FA is in charge of {\textcolor{black}{informing the teams of} the magnitudes and filters used to trigger the GRANDMA telescopes suitable for the follow-up to help characterize possible light curves. 
In the case of a counterpart discovery by GRANDMA telescopes, the FA {\textcolor{black}{had the responsibility} to trigger our spectroscopic facilities after checking if the transient magnitude was suitable for spectroscopy.
When the observation cycle for a particular GRB was complete, the FAs that received the alert during their shift reported the observations and their characteristics by sending out a GCN on behalf of GRANDMA.

The campaign presented in this paper was organized on 4 shifts per day: 04h-10h, 10h-16h, 16h-22h, and 22h-4h (UTC). 
The FAs were organized in weekly teams led by an experienced weekly coordinator (senior scientist) who was also responsible for training FAs, pairing the beginners with more experienced FAs, and supervising the full duration of the shift.
There are 4-8 shifters for a full week and a shifter will handle a particular slot once a day, starting and ending every Thursday (ending at about 16H30 Paris Time). 

We also trained the telescope teams to familiarize them with their role during the campaign and make sure they have access to all tools. 
As the major challenge of GRANDMA is to coordinate dozens of teams having different instruments all over the world, the training aimed to help bring all observational teams to the same level and perform ``smooth observations'' altogether.
The telescope teams are composed of at least one observer, and one photometry expert, that operates one (or more) telescopes, upon which several instruments can be mounted. 
{\textcolor{black}{In case of a transient alert, the FA on shift will warn the telescope teams about it.
The telescopes teams then must respond immediately whether or not they will be able to perform an observation of the given alert.
In a positive case, the teams will then provide images with a header containing standardized fields in as short a time as possible.}
The data will be then analyzed and their photometric results will be posted in the GCN. 
All the documentation available for telescope teams and follow-up adcates are available \href{https://grandma.ijclab.in2p3.fr/preparation-of-the-grb-campaign/}{at this page}.

\subsection{Photometry}
\label{datareduc}

The process of uniformly analyzing
the diverse set of images taken by different telescopes within our network was done by two dedicated photometric pipelines: {\sc STDPipe}\footnote{{{\sc STDPipe} is available at \url{https://gitlab.in2p3.fr/icare/stdpipe}}} 
\citep{stdpipe}
and {\sc Muphoten}\footnote{{{\sc MUphoten} is available at \url{https://gitlab.in2p3.fr/icare/MUPHOTEN}}} \citep{muphoten,2023arXiv230206225K}.
These software stacks were the same as those used during our previous ``ready for O4 campaign-I'' which is described in Section 4 of \citet{readyO4}.

These pipelines follow slightly different approaches. {\sc STDPipe} is a ready-to-use set of scripts pre-configured for processing the data from selected instruments. {\sc Muphoten} is a library of both low- and high-level routines for creation of custom pipelines for the data from arbitrary telescopes and varying complexity of the analysis (e.g. taking into account spatial dependence of photometric zero point or colour term, using custom noise models, advanced filtering of detected transient candidates, etc). In our analysis, they were configured to perform similar steps -- object detection, astrometric and photometric calibration, image subtraction using Pan-STARRS or telescope-specific templates, and transient detection and photometry on difference image -- on the data pre-processed by an instrument-specific code to perform bias, dark subtraction, and flat-fielding  on the telescope side. As shown in  \citet{readyO4}, the results of both pipelines are typically identical within measurement errors and provide an important cross-check on potential processing errors.


In the case of GRBs, the positions of the transient are known prior to our observations with the accuracy of, at least, 2-3 arcseconds due to promptly downlinked data from XRT. In case of events with optical afterglows, the position is further improved by the detections from Swift onboard Ultraviolet and Optical telescope (UT) or ground-based rapid response robotic telescopes, down to typically sub-arcsecond accuracy.
The latter was true for all the events where we had a clear detection of transients in our images. 
Therefore, we did not specifically perform any transient detection on our images; instead, we concentrated solely on the forced photometry at the published event position or different images, contingent upon the level of congestion in the field.
In cases where the object was not detectable, we derived the upper limits 
as described below.

In {\sc Muphoten}, they are computed as global properties of the whole studied image. The default method outlined in \citet{muphoten} calculates the success rate of recovering PS1 objects within 0.2\,mag intervals and selects the faintest interval where more than 10\% of PS1 objects in the field-of-view are detected in the image. In the case of images where there is a high detection rate up until the limit of the Pan-STARRS catalog, an alternative method defines the upper limit as the magnitude of the faintest source detected with an SNR of about 5.
For {\sc STDPipe}, we followed a simpler approach and assumed the 5$\sigma$ upper limit to be the flux equal to 5 times the background noise inside the aperture placed at the transient position.  It corresponds to the stellar magnitude of a point source that may be detected with a given SNR. 
Note that both these definitions are not sensitive to knowing exact transient position, and thus may be employed for placing the upper limits in cases of non-detection.

{\sc STDPipe} also contains a suite of routines that facilitate the proper detection and, to some extent, filtering of transients in the images, both before and after the template subtraction. While they have not been used during the GRB-focused run reported here, we plan to fully employ them for the task of transient detection in our data to be acquired during O4.

\section{Report summary of GRANDMA observations of GRBs}
\label{readyforO4}

In this section, we present the observation for the GRBs followed during the campaign. From 10 selected \textit{Swift} and one INTEGRAL GRB alerts, we successfully detected three afterglows. We highlight them in section 5.1: GRB\,220403B, GRB\,220427A and GRB\,220514A. Observations for two GRBs, GRB\,220404A and GRB\,220412B, were triggered but led to no observations as the targets were too close to the Sun and the Moon respectively. For six other GRBs, observations were made but no afterglow was detected by our telescopes, and their upper limits are reported in this section.

\subsection{Non-detections}

\textbf{GRB 220319A}. GRB 220319A was the first alert received in the ``2022 ReadyforO4 Campaign II'' on 19th March 2022 at 17:40:33 UT \citep{2022GCN.31769....1P}. The first observation occurred about 14 minutes \citep{2022GCN.31785....1B} after the initial \textit{Swift} alert thanks to amateur telescopes from the Kilonova Catcher program. 
Further follow-up was performed with ALi-50 telescope and nine amateur telescopes until $\sim$11 s after the GRB trigger, reporting a limiting magnitude of $\sim$20 ($B$,$R$ and Clear bands). None of these observations detected an afterglow: values consistent with observations made outside GRANDMA \citep{2022GCN.31770....1L,2022GCN.31773....1B,2022GCN.31774....1S,2022GCN.31775....1A,2022GCN.31776...1H}.

\textbf{GRB 220325A}. The \textit{Swift} Alert Telescope (BAT) detected the GRB 220325A on 25th March 2022, at 17:16:23 UT \citep{2022GCN.31787....1F}, and GRANDMA observations started 4.28 s after the BAT trigger time with SNOVA telescope \cite{2022GCN.31804....1A}. Further follow-up was obtained (in $r'$-, $R$- and $i$-band) with SNOVA, KAO, C2PU, and NOWT telescopes up to $\sim$10 s after the GRB trigger with a limiting magnitude up to 21.2\,mag, but no optical counterpart was detected \citep{2022GCN.31789....1L,2022GCN.31793....1G,2022GCN.31795....1S,2022GCN.31797....1N,2022GCN.31801....1K}.

\textbf{GRB 220408A}. GRB 220408A was detected by \textit{Swift} on 8th April 2022 at 05:46:04 UT and the first image was obtained $\sim 38$ minutes after the trigger from an amateur astronomer in $V$-band with a limiting magnitude of 17.7 without detection. Further observations were achieved with four GRANDMA telescopes, SRO, Xinglong-TNT, ALi-50, and MOSS, and two amateur astronomers (T19 and HAO) up to $\sim 19$ s after the \textit{Swift} trigger \citep{2022GCN.31884....1B}. None of these observations provided a detection of the afterglow with the most constraining limiting magnitude achieved being 21.4\,mag without a filter \citep{2022GCN.31855....1O,2022GCN.31859....1H, 2022GCN.31862....1S,2022GCN.31864....1Z,2022GCN.31870....1W}. 

\textbf{GRB 220412A}. On 12th April 2022 at 06:36:50 UT the \textit{Swift} triggered and located GRB 220412A which was classified as long with a duration of $T_{90}=41.66 \pm 7.00$ sec \citep{2022GCN.31881....1K}. GRANDMA observed GRB\,220412A with three telescopes and two amateur astronomers between $\sim$5.6 and $\sim$22 s after the \textit{Swift} trigger \citep{2022GCN.31903....1B}. The observations were contaminated by the moon and only provided upper limits in $L$, $R$, $V$, and $I$-band reporting a limiting magnitude up to $20.8$ in $R$-band \citep{2022GCN.31882....1W,2022GCN.31898....1W,2022GCN.31902....1W}.

\textbf{GRB 220430A}. GRB 220430A was detected on the 30th of April 2022 at 13:53:15 UT by \textit{Swift} \citep{2022GCN.31972....1A}.
GRANDMA observed GRB\,220430A with four telescopes, SNOVA, Makes-T60, HAO, and MOSS, between $\sim$34 minutes and $\sim$6.3 s after the \textit{Swift} trigger \citep{2022GCN.31977....1D}. None of these observations provided detection of the afterglow with the most constraining limiting magnitude achieved being 21.0\,mag without a filter \citep{2022GCN.31974....1G,2022GCN.31976....1P,2022GCN.31980....1H,2022GCN.31997....1S,2022GCN.31998....1J,2022GCN.31999....1J}.

\textbf{GRB 220501A}. On the 1st of May 2022 at 19:51:51 UT, the \textit{Swift} triggered and located GRB 220501A \citep{2022GCN.31982....1D}.
GRB\,220501A was observed by one amateur astronomer starting about $\sim$30 minutes after the trigger. We did not detect any optical counterpart to a limiting magnitude of 16.0 in $r$'-band \citep{2022GCN.31990....1H}.

\subsection{Afterglow detections}

\subsection*{GRB\,220403B}  

The \textit{Swift}-Burst Alert Telescope (BAT) observed and detected GRB 220403B (trigger=1101053), on 3rd April 2022, at 20:42:42 UT. BAT's light curve showed a single-peak structure with a duration of 30\,s, and a rate of about 2800\,counts/s (15-350 keV) \citep{2022GCN.31820....1K}.
The 1-second peak photon flux measured at 1.06 seconds in the 15-150 keV band was $2.7 \pm0.3$ ph/cm$^{2}$/sec: all the quoted errors are at the 90\% confidence level \citep{2022GCN.31834....1L}.

The first GRANDMA observation started eight minutes after the BAT alert \citep{2022GCN.31883....1S}. 
Follow-up was performed with four GRANDMA telescopes up to $\sim 1$ day after. 
The afterglow was clearly detected within the first  hour in $g$' and $r$'-band and observations without a filter.
The light curve without a filter can be fit by a power law decay with an index of $\alpha_O=0.41\pm0.07$ (where $f_\nu \propto t^{-\alpha} \nu^{-\beta}$). 
We fit, as shown in Figure \ref{fig:lc220403B}, the X-ray light curve observed by \textit{Swift}/XRT with a broken power law with two breaks. 
The obtained break time are $t_{b1}=263\pm2$s and $t_{b2}=9571\pm1667$s respectively with a decay index of $\alpha_{X1}=3.73\pm0.19$, $\alpha_{X2}=0.30\pm0.07$ and $\alpha_{X3}=1.16\pm0.08$ respectively. 
The optical data obtained fall in between the two breaks observed in X--rays and the shallow decay observed in optical is coincident with a plateau phase in X-ray. 
These plateaus are common in X-ray and optical and have already been observed in other GRBs (e.g. \citet{2017A&A...607A..84K}). They can be interpreted in a variety of ways: late energy injection in the forward shock due to a wide spread of the gamma factors (e.g. \citealt{Granot2006}), high latitude emission produced at large angles with respect to the jet axis leading to a long-lasting X-ray plateau for an on-axis observer (e.g. \citealt{Oganesyan2020,Ascenzi2020}), a structured jet seen slightly off-axis (e.g. \citealt{Beniamini2020}), the signature of a long-lived reverse shock due to a tail of low Lorentz factor material \citep[e.g.][]{uhm:07,genet:07,uhm:12}, evidence for a leftover GRB central engine being a millisecond magnetar (e.g. \citet{2019ApJS..245....1T,2020ApJ...896...42Z}), or emission due to continued accretion of fall-back gas onto the newborn neutron star or black hole (e.g. \citealt{2008Sci...321..376K, 2008MNRAS.388.1729K}). We note in the spectral domain that our color data point toward a spectral index $\beta_O \sim -1.4$. Albeit we have not corrected the data for optical extinction, all galactic extinction models predict a lower extinction in the r' compared to the g' band (e.g. \citet{Schlafly2011}). Thus, our measurement can be understood as an upper limit: $\beta_O < -1.4$. This value is highly unusual and not compatible with the expectations from the standard fireball model. Even if the injection frequency was located in the r' band, this would lead to a measure of $\beta_O \sim -0.3$ at best. The data we have in hand are however too scarce to investigate further this steep spectral index. No host galaxy is visible in our latest image with a limiting magnitude of 20.9 (in $B$ and $g$-band).

   \begin{figure}
   \centering
   \includegraphics[width=\hsize]{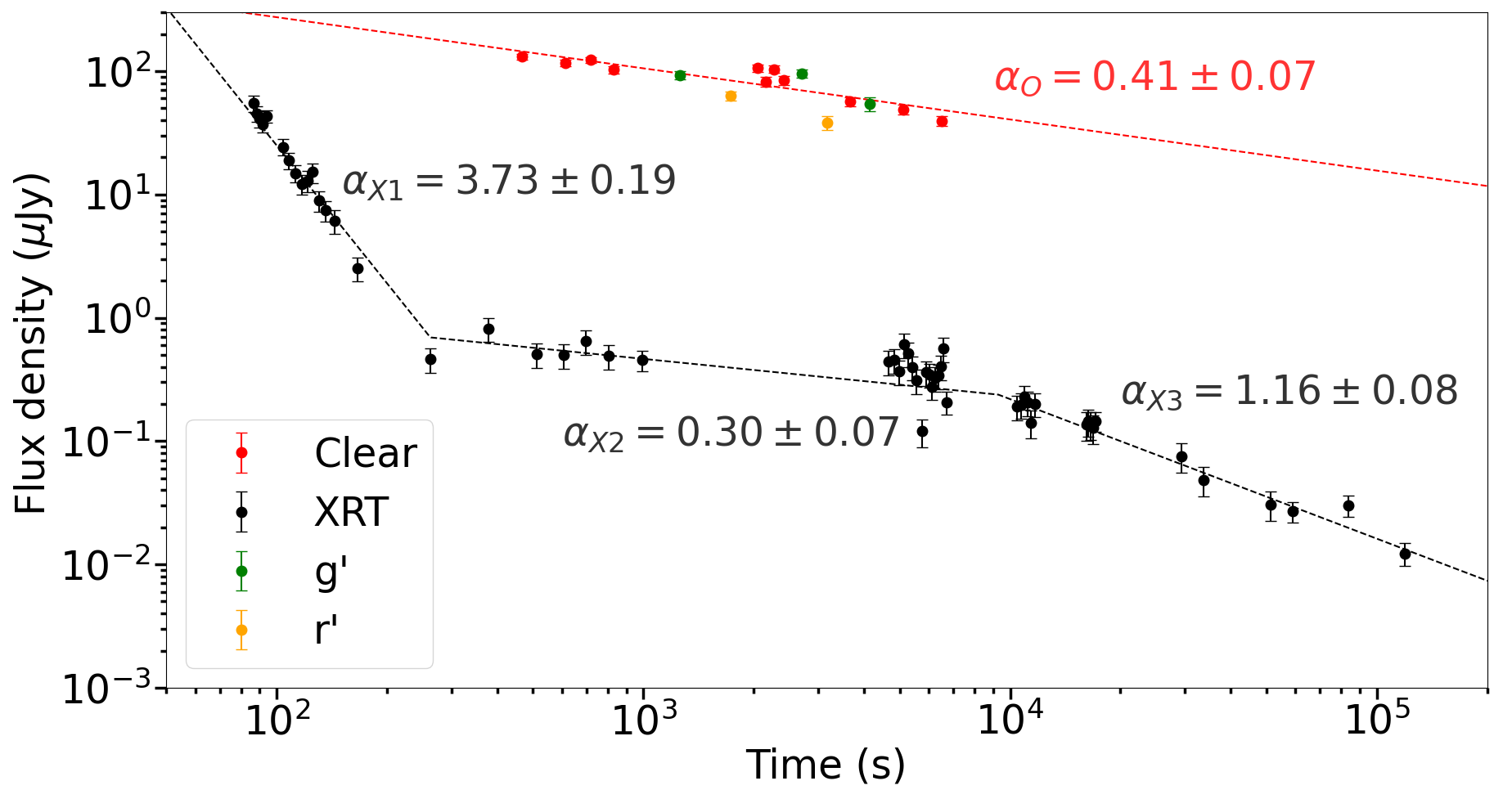}
      \caption{Optical and X-ray light curve of GRB\,220403B. The dashed line shows the best power law fit for both optical and X-ray light curves.
              }
         \label{fig:lc220403B}
   \end{figure}


\subsection*{GRB\,220427A}    

The GRANDMA system received the \textit{Swift} notice and transmitted the observation request to the whole network in about one minute. Early on, the GRB had limited observability from terrestrial observatories and was limited to observatories near Australia and the South of Africa. Within the GRANDMA network, only the telescopes located at La Réunion could observe at early times. The TAROT/TRE robotic telescope, automatically triggered by the alert, started the observation of the \textit{Swift} localization without a filter in the beam about 196\,s after the GRB trigger. Preliminary analysis of the TAROT/TRE images rapidly confirmed the detection of an optical afterglow. 
Further detection of the afterglow had been obtained by Les Makes/T60 telescope starting about 1.4\,hr after the GRB trigger.

The light curve elution derived from our optical data, presented in Figure \ref{fig:lc220427A}, can be fit by a power law decay with a decay index of $\alpha_O=1.38\pm0.04$. The X-ray light curve, as observed by \textit{Swift}/XRT, elves as a power law with a decay of $\alpha_X=1.08\pm0.03$. From X--rays, we also have the spectral slope, which corresponds to a $\beta_X=1.02+/-0.15$ (where $F=\nu^{-\beta}$). This corresponds to a standard fireball model with a wind density profile in the external medium, at a pre-break time, with the cooling break located between the optical and X-rays and an electron slope $p\simeq 2.2$ \citep{2002ApJ...568..820G}.

   \begin{figure}
   \centering
   \includegraphics[width=\hsize]{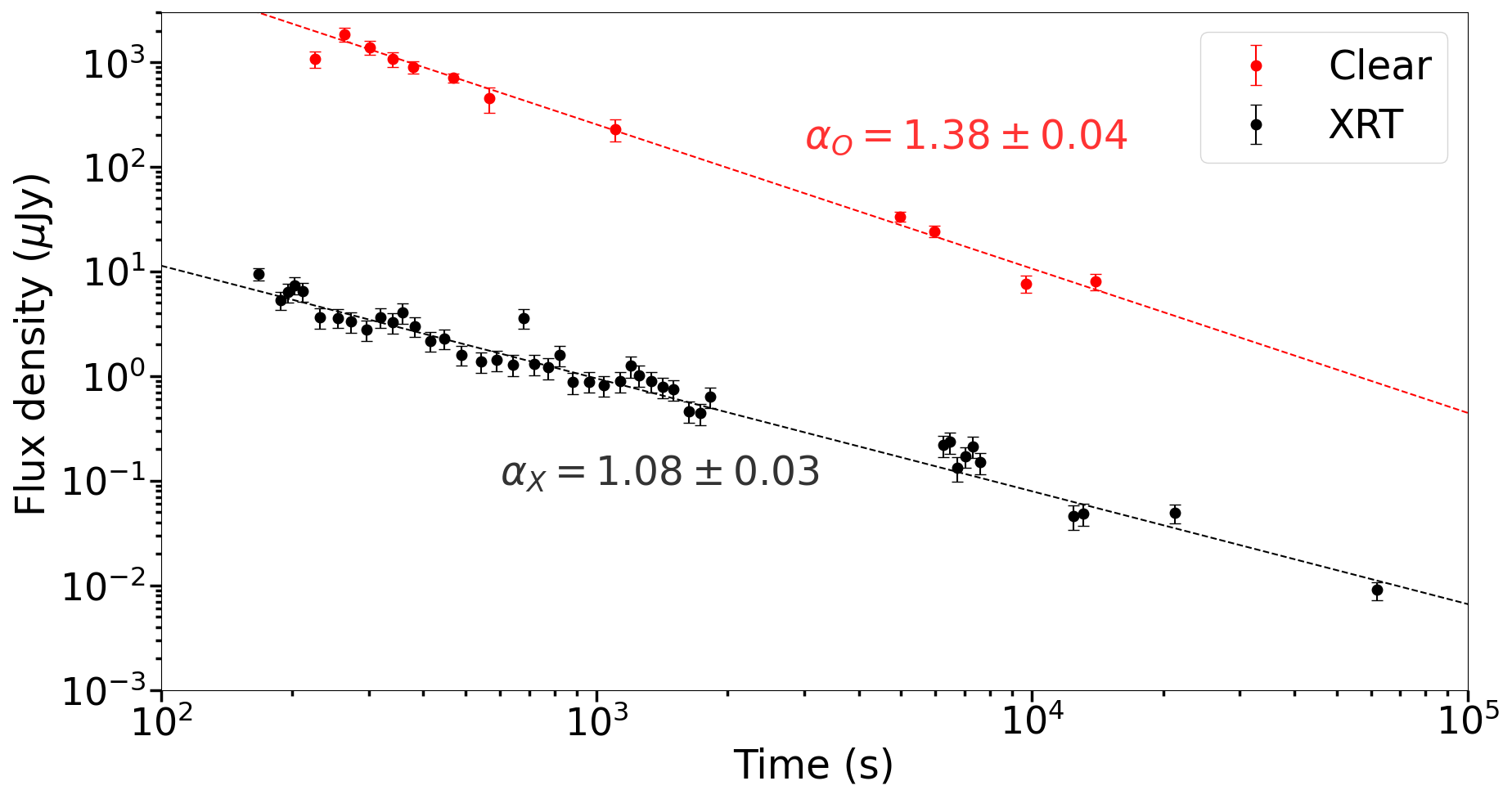}
      \caption{Optical and X-ray light curve of GRB\,220427A. The dashed line shows the best power law fit for both optical and X-ray light curves. 
              }
         \label{fig:lc220427A}
   \end{figure}

The host galaxy of GRB\,220427A was observed using GROND at the MPG 2.2\,m telescope at ESO's La Silla observatory \citep{Greiner2008} on 27 May 2022 in 7 bands ({\it grizJHK$_S$}). The observation included integration of $\sim$3600s in each of the bands. We corrected the resulting photometry of galactic extinction using the \citep{Schlegel1998} and the \citep{Schlafly2011} corrections.
The spectral energy distribution was used to determine a photometric redshift by fitting it with galaxy models at varying redshifts using LePhare \citep{Arnouts2011}. The best fit was obtained for a redshift of $z=0.82\pm0.09$ (see figure \ref{fig:host220427A}). Using this redshift, we performed a further SED fit with CIGALE \citep{Boquien2019}, which delivered similar host galaxy properties to the ones obtained with LePhare. The results obtained with CIGALE are given in Table \ref{Table:host220427A}. Input parameters for the SED fitting with CIGALE are presented in Table \ref{Table:paramshost220427A}. We have adopted a delayed star formation rate ($\rm SFR \propto t/\tau_0^2 \cdot e^{-t/\tau_0}$), on top of which a possible recent burst of star formation is enabled. The recent burst and the AGN contributions are not required to reproduce the obtained SED behavior and removing both contributions marginally changes the estimated parameters presented in Table \ref{Table:host220427A}. The inferred parameters are compatible with what one could expect from a long GRB host galaxy (e.g. \cite{2014A&A...565A.112H,2016SSRv..202..111P,2019A&A...623A..26P,2022A&A...666A..14S}).

   \begin{figure}
   \centering
   \includegraphics[width=\hsize]{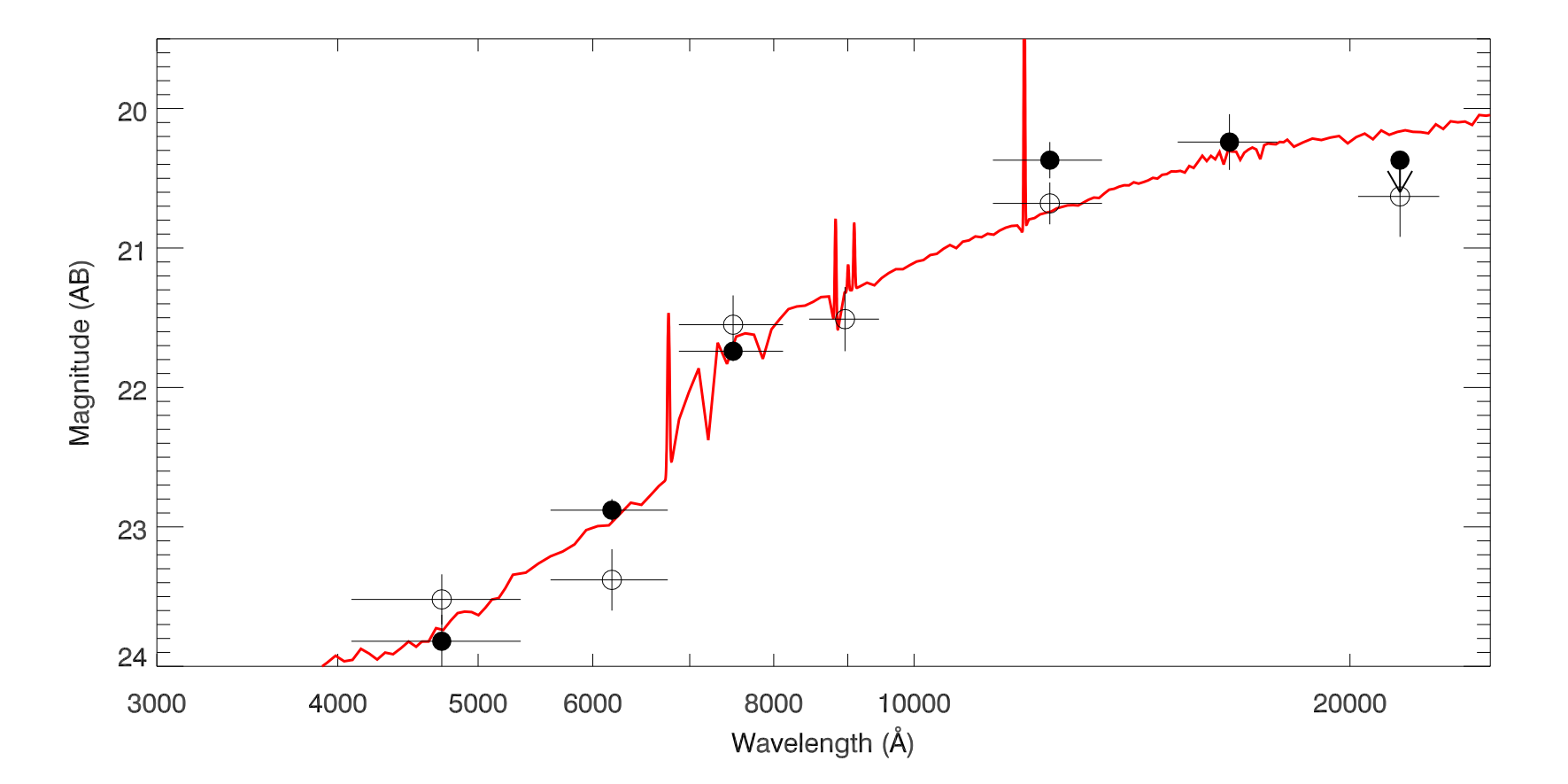}
      \caption{Spectral energy distribution fit to the host galaxy photometry of GRB\,220427A.
              }
         \label{fig:host220427A}
   \end{figure}

\subsection*{GRB\,220514A} 

GRB 220514A was detected by INTEGRAL \citep{2022GCN.32054....1B}. The preliminary analysis showed a light curve with multiple peaks in the $50-300$ keV range and a duration of around $66$ seconds, making it a long GRB. The on ground-calculated position for GRB220514A is RA $=147.6670$ deg, and Dec $+13.1472$ deg \citep{2022GCN.32054....1B}. 
From $T0-2.8s$ to $T0+64.8s$, the time-averaged spectrum was best fit using a band function with $Epeak = 139 \pm 21$ keV and an alpha value of $-1.25 \pm 0.06$. 
The 1-second peak photon flux starting at $T0+28.9$\,s was found to be in the $10-1000$ keV band range and is $16.6 \pm 0.5$  ph/s/$cm^2$ \citep{2022GCN.32054....1B}. 

The first observations of the GRANDMA telescope network were obtained by TNT $1.57$ h after trigger time, which detected an afterglow in $r$'-band with mag $18.9 \pm 0.1$, compatible with other telescopes reported outside of GRANDMA \citep{2022GCN.32070....1G, 2022GCN.32059....1M, 2022GCN.32051....1Z, 2022GCN.32044....1D, 2022GCN.32042....1L}. 
Further detections were made by MOSS and CAHA/CAFOS in Clear and $i$'-band respectively \citep{Yan2022GCN32058} (see Table \ref{tab:KN_observations}). Other observations by AbAO-T70, KNC-HAO, FRAM, and KNC-T21 were performed, resulting in upper limits. Due to the lack of multiple detections in any band, no slope could be fit for this GRB using GRANDMA data.

\section{Results and summary of the campaign}
\label{results}

This 8-week-long GRB follow-up campaign fulfilled its goal of training telescope teams for O4 as well as the improvement of the associated technical toolkits. 
One of the main goals was to stress the GRANDMA system in various ways, very similar to what is expected from O4 in order to simulate a real GW follow-up. 

In this way, GRANDMA's infrastructure has elved greatly since the first campaign \cite{GRANDMA_ztf_fink}. 
The GRANDMA system has been able to ingest and distribute the \textit{Swift} alerts to the network in real-time with a delay compatible with the detection of early afterglows. The online system harmonized the image collection and optimized the data reduction, which represents meticulous and time-consuming work for this kind of campaign, covering numerous observations with very different telescopes. This campaign further allowed testing and standardizing of the two data analysis pipelines (\texttt{STDpipe} and \texttt{MUphoten}) used by the collaboration.

As presented in Figure \ref{fig:piechart}, of the 11 triggers selected, nine fields have been observed, and three afterglows were detected. The network was not able to perform any observations for two of these 11 GRBs due to the localization of the source near the Sun and/or the Moon. For seven of them, the first observation occurred less than one hour after the BAT trigger. The fastest image achieved thanks to the TAROT robotic telescope system, was performed 196\,s after the GRB\,220427A BAT trigger. On the KNC side, the fastest observation was performed 14 minutes after the GRB\,220319A BAT trigger, highlighting the effectiveness of the KNC system, the willingness of amateurs to contribute, and their ability to provide scientifically valuable data. 
A total of 17 professional telescopes and 17 amateur astronomers' participated in the "Ready for 04 II campaign", providing good-quality images from which upper limits were computed (see Table \ref{tab:selected_observations}).


The time sampling of the observation of GRB\,220427A allowed us to study the lightcurve elution and fit the decay slope of the afterglow. The long-term follow-up allowed us to identify and study the properties of the host galaxy which exhibits typical properties for a long GRB host galaxy.
The efficiency of the network to observe GRB afterglow is \textcolor{black}{illustrated} in Figure \ref{fig:afterglowefficiency}. 
It presents a selection of one observation per telescope (gathered in Table \ref{tab:selected_observations}) that was \textcolor{black}{performed} during the campaign, and compares its delay and limiting magnitude to an \textcolor{black}{archived} sample of an observed afterglow lightcurve. Figure \ref{fig:afterglowefficiency} highlights the performances of the observation performed during the campaign. \textcolor{black}{All selected observations would have been able to detect a significant fraction of optically bright archived afterglows in the sample.}


\section{Conclusions}
\label{conclusions}

In this paper, we describe GRANDMA's ''Ready for O4 II'', a campaign designed to train the network of telescopes as well as to test new tools for MM purposes.
''Ready for O4 II'' observed and searched for afterglows of \textit{Swift} GRBs during an 8-week period.
This campaign was essential to improve GRANDMA's network and system to be ready for the IGWN O4 run, as the follow-up of GW alerts is very challenging and preparing our collaboration will greatly enhance the chances of efficiently observing an electromagnetic counterpart during future follow-ups.

   \begin{figure}
   \centering
   \includegraphics[width=\hsize]{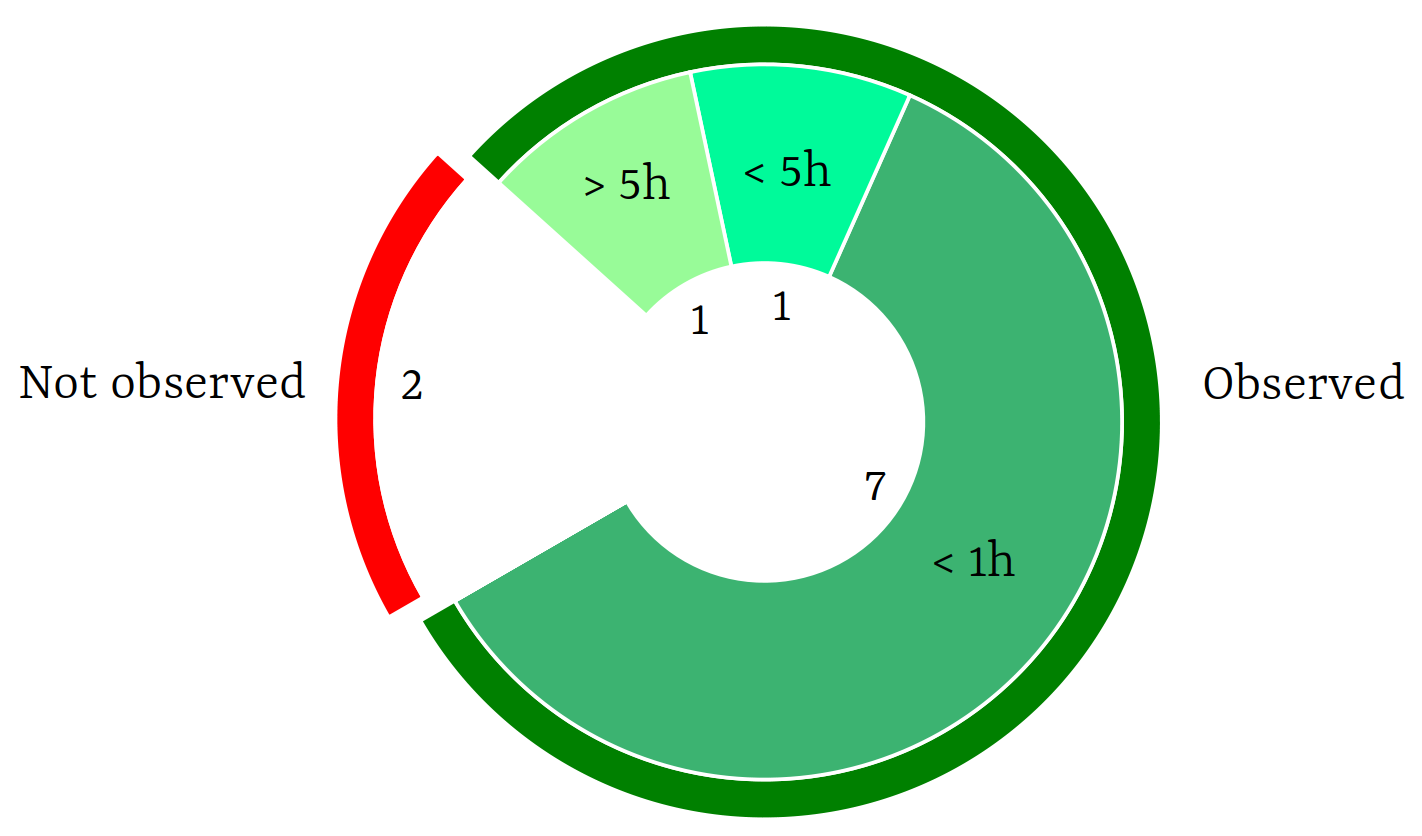}
      \caption{Overall performance of the GRB follow-up conducted by GRANDMA telescopes. }
         \label{fig:piechart}
   \end{figure}

   \begin{figure}
   \centering
   \includegraphics[width=\hsize]{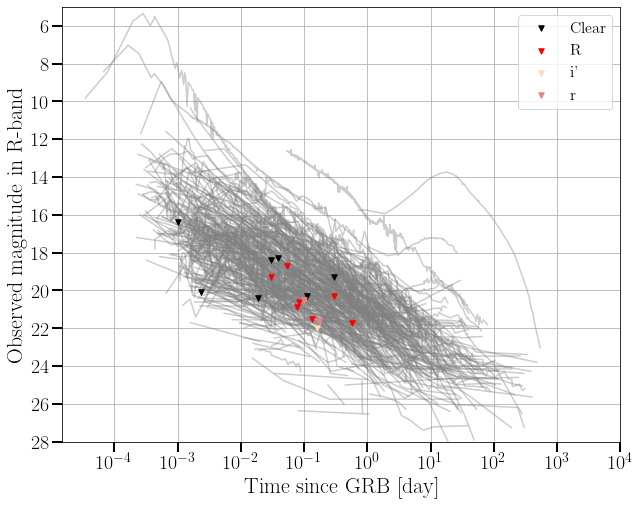}
      \caption{Selected \textcolor{black}{achieved} upper limits of observation \textcolor{black}{performed} during the campaign (gathered in Table \ref{tab:selected_observations}) compared to a sample of observed afterglow lightcurve in R band.}
         \label{fig:afterglowefficiency}
   \end{figure}

Here, we stressed the challenge of coordinating such a large network as GRANDMA both in terms of infrastructure and human resources.  We presented developments performed at the interface and communication levels as well as in the image collection, processing, and data reduction to uniformize and authorize the follow-up results. 
This also provided important practice for the collaboration to become accustomed to using the reduction pipelines, {\sc STDpipe} and {\sc MUphoten}.
From the follow-up of 11 selected \textit{Swift}/INTEGRAL GRB alerts, we successfully observed nine GRBs, representing 82\% of the total triggers, and detected three afterglows (GRB\,220403B, GRB\,220427A, GRB\,220514A), in an 8-week-long campaign.  

{\textcolor{black}{Our response to the GRB alerts was less than an hour for eight of them, meaning that our teams were able to start observations within less than one hour after the initial time of the trigger.} 
For GRB\,220427A, we presented the detection of the host galaxy and the study of its properties including its redshift. Altogether, the observations presented in this work show the capacity of the network to contribute meaningfully to multi-messenger follow-up and the importance of this work to tackle the challenges of the transient sky in the coming years.

\begin{acknowledgements}

GRANDMA Collaboration thanks sincerely the participation and wise advice of Dr. Kann Alex who was a remarkable scientist, and his untimely death has left us with big gaps. Dr. Kann's contribution to science will live on through his work. 
I.T.M. is supported by projects financed by PNRR measures with European Union funds – NextGenerationEU. F.D., J.-G.D. and C.P. acknowledge financial support from the Centre National d’Études Spatiales (CNES).
J.-G.D. is supported by a research grant from the Ile-de-France Region within the framework of the Domaine d’Intérêt Majeur-Astrophysique et Conditions d’Apparition de la Vie (DIM-ACAV). The Kilonova-Catcher program is supported by the IdEx Universit\'e de Paris Cit\'e, ANR-18-IDEX-0001. This project has received financial support from the CNRS through the MITI interdisciplinary programs. SA is supported by the CSI-Recherche Université Côte d'Azur. MWC is supported by the National Science Foundation with grant numbers PHY-2308862 and OAC-2117997. AT, AF, and AS acknowledge support from Science, Technology \& Innovation Funding Authority (STDF) under grant number 45779. Part of the funding for GROND (both hardware as well as personnel) was generously granted from the Leibniz-Prize to Prof. G. Hasinger (DFG grant HA 1850/28-1).
SK is supported by European Structural and Investment Fund and the Czech Ministry of Education, Youth and Sports (Project CoGraDS -- CZ.02.1.01/0.0/0.0/15\_003/0000437).
The work of FN is supported by NOIRLab, which is managed by the Association of Universities for Research in Astronomy (AURA) under a cooperative agreement with the National Science Foundation. 
Development of data analysis techniques and {\sc STDPipe} package, as well as follow-up adcate work of MM were partially supported by the grants of the Ministry of Education of the Czech Republic LM2023032 and LM2023047. J. M. is supported by the National Key R\&D Program of China (2023YFE0101200), Yunnan Revitalization Talent Support Program (YunLing Scholar Award), and NSFC 11673062. W.~Corradi and N.Sasaki thank the support from the Brazilian Funding Agencies Capes, CNPq, Fapemig and our home institutions UFMG and UEA.
The work of X. W. is supported by the National Science Foundation of China (NSFC grants 12288102, 12033003, 11633002 and 12090044), the Ma Huateng Foundation, the Scholar Program of Beijing Academy of Science and Technology (DZ:BS202002), and the Tencent Xplorer Prize. The work of SNOVA is supported by the High-Level Talent–Heaven Lake Program of Xinjiang Uygur Autonomous Region of China, the National Natural Science Foundation of China (NSFC, grants 11803076, 12203029)), and the National Key R\&D program of China for Intergovernmental Scientific and Technological Innovation Cooperation Project under No. 2022YFE0126200.

\end{acknowledgements}

\onecolumn
\begin{table}
	\renewcommand{\arraystretch}{1.4}
	\centering
    \label{Table:host220427A}
	\caption{Summary of SED fitting results. (1) stellar mass in log scale, (2) SFR in log scale, (3) amount of dust attenuation in the V band, (4) metallicity, (5) age weighted by stellar mass, (6) attenuation curve slope for stellar continuum, (7) AGN fraction, (8) mass fraction of the late burst population.} 

    \begin{tabular}{cccccccc}
    \\ \hline

    log$_{10}$(M$_{\star}$) & log$_{10}$(SFR) & A$_V$ & $Z$ & t$_{mass}$ & $\delta$ & $f_{agn}$ & $f_{burst}$ \\
    
	[M$_{\sun}$] & [M$_{\sun}$ yr$^{-1}$] & mag & & Myrs & & & \\ \hline
$10.62_{-0.32}^{+0.18}$ & $1.89_{-nan}^{+0.31}$ & $1.32\pm0.51$ & $0.004_{-0.004}^{+0.007}$ & $802\pm668$ & $-0.60\pm0.30$ & $0.03_{-0.03}^{+0.06}$ & $0.08_{-0.08}^{+0.08}$ \\ \hline

    \end{tabular}
\end{table}
\begin{table}
	\centering
	\caption{Input parameters for SED fitting with CIGALE.}
	\label{Table:paramshost220427A}
	\begin{tabular}{c c c}
	\hline
	Parameter & Symbol & Range \\
	\hline\hline
	{\textbf{Delayed star formation history $+$ recent burst}}\\
	
	Age of the main stellar population & $t_0$ &  200, 500, 1000, 2000, 3000 \,Myrs \\
	E-folding time of the main stellar population model & $\tau$ &  200, 500, 1000, 2000, 3000, 5000, 7000, 10000  Myrs\\
	Mass fraction of the late burst population & $f_{burst}$ &  0.0, 0.1, 0.2\\
	Age of the late burst & $t_b$ &  20.0, 50.0, 100.0, 200.0\\
	E-folding time of the late starburst population & $\tau_{b}$ &  50.0\\
	\hline
	Metallicity $^1$ & $Z$ & 0.0001, 0.008, 0.02\\
    \hline
	{\textbf{Dust attenuation}}\\
	Color excess for nebular emission $^2$ & $\rm E(B-V)_{lines}$ & 0.01, 0.25, 0.5, 0.75, 1.0, 1.25, 1.5, 1.75, 2.0, 2.25, 2.5, 2.75, 3.0 \\
	Ratio of color excess & $\cfrac{E(B-V)_{stars}}{E(B-V)_{lines}}$ & 0.44 \\
	Attenuation curve slope for stellar continuum & $\delta$ & -1, -0.7, -0.4, -0.1, 0.2, 0.5, 0.7\\
	\hline
	{\textbf{AGN$^3$}}\\
	Slope of the power law & $\alpha$ & 2.0\\
	AGN fraction & $f_{agn}$ & 0.0, 0.1, 0.2 \\
	\hline
	Redshift & $z$ & 0.8159 \\
	\hline	
	\hline
	\end{tabular}
	\begin{flushleft}
      \small
      \item $^1$ {$Z$=0.02 being the solar metallicity}
      \item $^2$ {$\rm E(B-V)_{lines}$ is the color excess between the B and V bands applied on the nebular emission lines.}
      \item $^3$ {Parameters of the dale2014 model \cite{Boquien2019}.}
    \end{flushleft}

\end{table}
\twocolumn

%
%

\section*{Affiliations}
$^{1}$ Department of Physics and Astronomy, University of Catania, 95125 Catania, Italy. e-mail: iara.tosta.melo@dfa.unict.it\\ 
$^{2}$ Sorbonne Universit\'e, CNRS, UMR 7095, Institut d’Astrophysique de Paris, 98 bis bd Arago, 75014 Paris, France. e-mail: ducoin@iap.fr \\ 
$^{3}$ N.Tusi Shamakhy Astrophysical Observatory Azerbaijan National Academy of Sciences, settl.Y. Mammadaliyev, AZ 5626, Shamakhy, Azerbaijan\\ 
$^{4}$ School of Physics and Astronomy, University of Minnesota, Minneapolis, Minnesota 55455, USA\\ 
$^{5}$ Brown University, Providence, RI 02912, Rhode Island, USA\\ 
$^{6}$ Oukaimeden Observatory,High Energy Physics and Astrophysics Laboratory, FSSM, Cadi Ayyad University  Av. Prince My Abdellah, BP 2390 Marrakesh, Morocco\\ 
$^{7}$ Universit\'e C\^ote d'Azur, Observatoire de la C\^ote d'Azur, CNRS, Laboratoire J.-L. Lagrange, Boulevard de l’Observatoire, 06304 Nice, France\\ 
$^{8}$ 14 rue Saint-Hubert, F-60560 Orry-la-Ville, France\\ 
$^{9}$ E. Kharadze Georgian National Astrophysical Observatory, Mt.Kanobili, Abastumani, 0301, Adigeni, Georgia\\ 
$^{10}$ Samtskhe-Javakheti  State  University, Rustaveli Str. 113,  Akhaltsikhe, 0080,  Georgia\\ 
$^{11}$ N.Tusi Shamakhy astrophysical Observatory Azerbaijan National Academy of Sciences, settl.Mamedaliyev, AZ 5626, Shamakhy, Azerbaijan\\ 
$^{12}$ Universit\'e C\^ote d'Azur, Observatoire de la C\^ote d'Azur, CNRS, Artemis, Boulevard de l’Observatoire, 06304 Nice, France\\ 
$^{13}$ YNAO\\ 
$^{14}$ Astronomical Observatory Taras Shevshenko National University of Kyiv, Observatorna str. 3, Kyiv, 04053, Ukraine\\ 
$^{15}$ Silesian University of Technology, Faculty of Automatic Control, Electronics and Computer Science, Akademicka 16, 44-100 Gliwice, Poland\\ 
$^{16}$ Astrophysique Relativiste Théories Expériences Métrologie Instrumentation Signaux, Nice, France\\ 
$^{17}$ GRAPPA, Anton Pannekoek Institute for Astronomy and Institute of High-Energy Physics, University of Amsterdam, Science Park 904,1098 XH Amsterdam, The Netherlands\\ 
$^{18}$ FZU - Institute of Physics of the Czech Academy of Sciences, Na Slovance 1999/2, CZ-182 21, Praha, Czech Republic\\ 
$^{19}$ Vereniging or Sterrenkunde, Balen-Neetlaan 18A, 2400, Mol, Belgium\\ 
$^{20}$ Ulugh Beg Astronomical Institute, Uzbekistan Academy of Sciences, Astronomy str. 33, Tashkent 100052, Uzbekistan\\ 
$^{21}$ Centre for Astrophysics and Supercomputing, Swinburne University of Technology, Hawthorn, Victoria 3122, Australia ARC Centre of Excellence for Gravitational Wave Discovery {\it (OzGrav)}\\ 
$^{22}$ Laborat\'orio Nacional de Astrof\'isica, R. dos Estados Unidos, 154 - Na\c{c}\~oes, Itajub\'a - MG, 37504-364, Brazil\\ 
$^{23}$ Institut Universitaire de France, Minist\`ere de l’Enseignement Sup\'erieur et de la Recherche, 75231 Paris, France\\ 
$^{24}$ CPPM, Aix Marseille Univ, CNRS/IN2P3, CPPM, Marseille, France\\ 
$^{25}$ Universit\'e Paris Cit\'e, CNRS, Astroparticule et Cosmologie, F-75013 Paris, France\\ 
$^{26}$ National University of Uzbekistan, 4 University str., Tashkent 100174, Uzbekistan\\ 
$^{27}$  \\ 
$^{28}$ National Research Institute of Astronomy and Geophysics (NRIAG), 1 El-marsad St., 11421 Helwan, Cairo, Egypt\\ 
$^{29}$ KNC, AAVSO, Hidden Valley Observatory(H), Colfax, WI.; iTelescope, NMS, Mayhill, NM.\\ 
$^{30}$ Department of Physics, University of Western Australia, Crawley WA 6009, Australia\\ 
$^{31}$ Australia ARC Centre of Excellence for Gravitational Wave Discovery {\it (OzGrav)}\\ 
$^{32}$ Physics Department and Astronomy Department, Tsinghua University, Beijing, 100084, People's Republic of China\\ 
$^{33}$ Department of Aerospace, Physics, and Space Sciences, Florida Institute of Technology, Melbourne, Florida 32901, USA\\ 
$^{34}$ American University of Sharjah, Physics Department, PO Box 26666, Sharjah, UAE\\ 
$^{35}$ Institut f\"ur Physik und Astronomie, Universit\"at Potsdam, Karl-Liebknecht-Str. 24/25, D-14476 Potsdam, Germany\\ 
$^{36}$ IJCLab, Univ Paris-Saclay, CNRS/IN2P3, Orsay, France\\ 
$^{37}$ Samtskhe-Javakheti State University, Rustaveli Str. 113, Akhaltsikhe, 0080, Georgia\\ 
$^{38}$ Xinjiang Astronomical Observatory, Chinese Academy of Sciences, Urumqi, Xinjiang 830011, People's Republic of China\\ 
$^{39}$ School of Astronomy and Space Science, University of Chinese Academy of Sciences, Beijing 100049, People's Republic of China\\ 
$^{40}$ Oukaimeden Observatory HAO telescope, Oukaimeden, Morocco.\\ 
$^{41}$ Division of Physics, Mathematics, and Astronomy, California Institute of Technology, Pasadena, CA 91125, USA\\ 
$^{42}$ Hessian Research Cluster ELEMENTS, Giersch Science Center, Max-n-Laue-Stra{\ss}e 12, Goethe University Frankfurt, Campus Riedberg, 60438 Frankfurt am Main, Germany\\ 
$^{43}$ CEICO, Institute of Physics of the Czech Academy of Sciences, Na Slovance 1999/2, CZ-182 21, Praha, Czech Republic\\ 
$^{44}$ Laboratoire de Physique et de Chimie de l'Environnement, Universit\'e Joseph KI-ZERBO, Ouagadougou, Burkina Faso\\ 
$^{45}$ IRAP, Universit\'e de Toulouse, CNRS, UPS, 14 Avenue Edouard Belin, F-31400 Toulouse, France\\ 
$^{46}$ Universit\'e Paul Sabatier Toulouse III, Universit'e de Toulouse, 118 route de Narbonne, 31400 Toulouse, France\\ 
$^{47}$ K26 / Contern Observatory (private obs.), 1, beim Schmilberbour, 5316 Contern, Luxembourg\\ 
$^{48}$ Centre for Cosmology, Particle Physics and Phenomenology - CP3, Universite Catholique de Louvain, B-1348 Louvain-la-Neuve, Belgium\\ 
$^{49}$ Physics Department and Astronomy Department, Tsinghua University, Beijing, 100084, People's Republic of China; Beijing Planetarium, Beijing Academy of Science and Technology, Beijing, 100044, People's Republic of China\\ 
$^{50}$ Yunnan Observatories, Chinese Academy of Sciences, Kunming 650011, Yunnan Province, People's Republic of China\\ 
$^{51}$ Key Laboratory for the Structure and Elution of Celestial Objects, Chinese Academy of Sciences, 650011 Kunming, People's Republic of China\\ 
$^{52}$ Observatoire du Crous des Gats, F-31550 Cintegabelle, France\\ 
$^{53}$ Soci\'t\' astronomique de France, 3 rue Beethoven, F-75016 Paris, France\\ 
$^{54}$ University of the Virgin Islands, United States Virgin Islands 00802, USA\\ 
$^{55}$ E. Kharadze Georgian National Astrophysical Observatory, Mt.eonobili, Abastumani, 0301, Adigeni, Georgia\\ 
$^{56}$ SOAR Telescope/NSF's NOIRLab, Avda Juan Cisternas 1500, 1700000, La Serena, Chile\\ 
$^{57}$ Th\"uringer Landessternwarte Tautenburg, 07778 Tautenburg, Germany\\ 
$^{58}$ National Astronomical Research Institute of Thailand (Public Organization), 260, Moo 4, T. Donkaew, A. Mae Rim, Chiang Mai, 50180, Thailand\\ 
$^{59}$ Hankasalmi Observatory, Jyvaskylan Sirius ry, Verkkoniementie 30, FI-40950 Muurame, Finland\\ 
$^{60}$ OrangeWave Innovative Science, LLC, Moncks Corner, SC 29461, USA\\ 
$^{61}$ Artemis, Observatoire de la Côte d’Azur, Université Côte d’Azur, Boulevard de l'Observatoire, 06304 Nice, France\\ 
$^{62}$ Silesian University of Technology, Department of Electronics, Electrical Engineering and Microelectronics, Akademicka 16, 44-100 Gliwice, Poland\\ 
$^{63}$ Universit\'e de Strasbourg, CNRS, IPHC UMR 7178, F-67000 Strasbourg, France\\ 
$^{64}$ Astronomical Observatory of Taras Shevchenko National University of Kyiv, Observatorna Str. 3, Kyiv, 04053, Ukraine\\ 
$^{65}$ Max-Planck-Institut f\"ur extraterrestrische Physik, Gie{\ss}enbachstra{\ss}e 1, 85748 Garching, Germany\\ 
$^{66}$ Oukaimeden Observatory (MOSS), Oukaimeden,  Morocco\\ 
$^{67}$ Universit\'e Paris-Saclay, Universit\'e Paris Cit\'e, CEA, CNRS, AIM, 91191, Gif-sur-Yvette, France\\ 
$^{68}$ Soci\'et\'e Astronomique de France, Observatoire de Dauban, FR 04150 Banon, France\\ 
$^{69}$ Astronomy and Space Physics Department, Taras Shevchenko National University of Kyiv, Glushkova ave., 4, Kyiv, 03022, Ukraine\\ 
$^{70}$ National Center Junior academy of sciences of Ukraine, 38-44, Dehtiarivska St., Kyiv, 04119, Ukraine\\ 
$^{71}$ Astronomical Observatory\ Taras Shevshenko National University of Kyiv, Observatorna str. 3, Kyiv, 04053, Ukraine\\ 
$^{72}$ Main Astronomical Observatory of National Academy of Sciences of Ukraine, 27 Acad. Zabolotnoho Str., Kyiv, 03143, Ukraine\\ 
$^{73}$ Beijing Planetarium, Beijing Academy of Science and Technology, Beijing, 100044, People's Republic of China\\ 
$^{74}$ Astronomy and Space Physics Department, Taras Shevchenko National University of Kyiv, Glushkova Ave., 4, Kyiv, 03022, Ukraine\\ 
$^{75}$ National Center Junior Academy of Sciences of Ukraine, Dehtiarivska St. 38-44, Kyiv, 04119, Ukraine\\ 
$^{76}$ Observatoire de la C\^ote d'Azur, Universit\'e C\^ote d'Azur, CNRS, UMS Galil\'ee, France\\ 
$^{77}$ Xinjiang Astronomical Observatory, 150 Science 1-Street, Urumqi, Xinjiang 830011, China\\ 
$^{78}$ KNC Deep Sky Chile Observatory\\ 
$^{79}$ Beijing Planetarium, Beijing Academy of Sciences and Technology, Beijing, 100044, China\\ 
$^{80}$ Physics department and Tsinghua Center for Astrophysics, Tsinghua University, Beijing, 100084, China\\ 
$^{81}$ Key Laboratory of Optical Astronomy, National Astronomical Observatories, Chinese Academy of Sciences, A20, Datun Road, Chaoyang District, Beijing 100012, People's Republic of China

\bibliographystyle{aa}
\bibliography{references}

\begin{thebibliography}{108}
\expandafter\ifx\csname natexlab\endcsname\relax\def\natexlab#1{#1}\fi

\bibitem[{{Abbott} {et~al.}(2017{\natexlab{a}}){Abbott}, {Abbott}, {Abbott},
  {Acernese}, {Ackley}, {Adams}, {Adams}, {Addesso}, {Adhikari}, {Adya}, \&
  {Affeldt}}]{LSC_BNS_2017PhRvL}
{Abbott}, B.~P., {Abbott}, R., {Abbott}, T.~D., {et~al.} 2017{\natexlab{a}},
  \prl, 119, 161101

\bibitem[{{Abbott} {et~al.}(2017{\natexlab{b}}){Abbott}, {Abbott}, {Abbott},
  {Acernese}, {Ackley}, {Adams}, {Adams}, {Addesso}, {Adhikari}, {Adya}, \&
  {Affeldt}}]{LSC_MM_2017ApJ}
{Abbott}, B.~P., {Abbott}, R., {Abbott}, T.~D., {et~al.} 2017{\natexlab{b}},
  \apjl, 848, L12

\bibitem[{Aivazyan {et~al.}(2022)Aivazyan, Almualla, Antier, Baransky,
  Barynova, Basa, Bayard, Beradze, Berezin, Blazek, Boutigny, Boust, Broens,
  Burkhonov, Cailleau, Christensen, Cejudo, Coleiro, Coughlin, Datashvili,
  Dietrich, Dolon, Ducoin, Duverne, Marchal-Duval, Galdies, Granier, Godunova,
  Gokuldass, Eggenstein, Freeberg, Hello, Inasaridze, Ishida, Jaquiery, Kann,
  Kapanadze, Karpov, Kiendrebeogo, Klotz, Kneip, Kochiashvili, Kou, Kugel,
  Lachaud, Leonini, Leroy, Leroy, Van~Su, Marchais, Masek, Midavaine, Moller,
  Morris, Natsvlishvili, Navarete, Noysena, Nissanke, Noonan, Orange, Peloton,
  Popowicz, Pradier, Prouza, Raaijmakers, Rajabov, Richmond, Romanyuk,
  Rousselot, Sadibekova, Serrau, Sokoliuk, Song, Simon, Stachie, Taylor,
  Tillayev, Turpin, Vardosanidze, Vlieghe, Melo, Wang, \&
  Zhu}]{GRANDMA_ztf_fink}
Aivazyan, V., Almualla, M., Antier, S., {et~al.} 2022, GRANDMA Observations of
  ZTF/Fink Transients during Summer 2021

\bibitem[{{Aivazyan} {et~al.}(2022){Aivazyan}, {Almualla}, {Antier},
  {Baransky}, {Barynova}, {Basa}, {Bayard}, {Beradze}, {Berezin}, {Blazek},
  {Boutigny}, {Boust}, {Broens}, {Burkhonov}, {Cailleau}, {Christensen},
  {Cejudo}, {Coleiro}, {Coughlin}, {Datashvili}, {Dietrich}, {Dolon}, {Ducoin},
  {Duverne}, {Marchal-Duval}, {Galdies}, {Granier}, {Godunova}, {Gokuldass},
  {Eggenstein}, {Freeberg}, {Hello}, {Inasaridze}, {Ishida}, {Jaquiery},
  {Kann}, {Kapanadze}, {Karpov}, {Kiendrebeogo}, {Klotz}, {Kneip},
  {Kochiashvili}, {Kou}, {Kugel}, {Lachaud}, {Leonini}, {Leroy}, {Leroy}, {Le
  Van Su}, {Marchais}, {Masek}, {Midavaine}, {Moller}, {Morris},
  {Natsvlishvili}, {Navarete}, {Noysena}, {Nissanke}, {Noonan}, {Orange},
  {Peloton}, {Popowicz}, {Pradier}, {Prouza}, {Raaijmakers}, {Rajabov},
  {Richmond}, {Romanyuk}, {Rousselot}, {Sadibekova}, {Serrau}, {Sokoliuk},
  {Song}, {Simon}, {Stachie}, {Taylor}, {Tillayev}, {Turpin}, {Vardosanidze},
  {Vlieghe}, {Tosta e Melo}, {Wang}, \& {Zhu}}]{readyO4}
{Aivazyan}, V., {Almualla}, M., {Antier}, S., {et~al.} 2022, arXiv e-prints,
  arXiv:2202.09766

\bibitem[{{Alexander} {et~al.}(2017){Alexander}, {Berger}, {Fong}, {Williams},
  {Guidorzi}, {Margutti}, {Metzger}, {Annis}, {Blanchard}, {Brout}, {Brown},
  {Chen}, {Chornock}, {Cowperthwaite}, {Drout}, {Eftekhari}, {Frieman}, {Holz},
  {Nicholl}, {Rest}, {Sako}, {Soares-Santos}, \&
  {Villar}}]{2017ApJ...848L..21A}
{Alexander}, K.~D., {Berger}, E., {Fong}, W., {et~al.} 2017, \apjl, 848, L21

\bibitem[{{Ambrosi} {et~al.}(2022){Ambrosi}, {D'Ai}, {D'Elia}, {Gronwall},
  {Gropp}, {Kennea}, {Lien}, {Marshall}, {Page}, {Palmer}, {Tohuvavohu}, \&
  {Neil Gehrels Swift Observatory Team}}]{2022GCN.31972....1A}
{Ambrosi}, E., {D'Ai}, A., {D'Elia}, V., {et~al.} 2022, GRB Coordinates
  Network, 31972, 1

\bibitem[{{Andrade} {et~al.}(2022){Andrade}, {Coughlin}, {Noysena},
  {Kiendrebeogo}, {Melo}, {Wang}, {Navarete}, {Zhu}, {Wang}, {Iskandar},
  {Zeng}, {Abe}, {Bendjoya}, {Rivet}, {Vernet}, {Antier}, {de Ugarte Postigo},
  {Fouad}, {Takey}, {Shokry}, {Elhosseiny}, {Eid}, {Farouk}, {Duverne}, \&
  {GRANDMA Collaboration}}]{2022GCN.31804....1A}
{Andrade}, C., {Coughlin}, M., {Noysena}, K., {et~al.} 2022, GRB Coordinates
  Network, 31804, 1

\bibitem[{{Andreoni} {et~al.}(2017){Andreoni}, {Ackley}, {Cooke},
  {et~al.}}]{Andreoni_2017PASA}
{Andreoni}, I., {Ackley}, K., {Cooke}, J., {et~al.} 2017, PASA, 34, e069

\bibitem[{{Antier} {et~al.}(2020{\natexlab{a}}){Antier}, {Agayeva}, {Aivazyan},
  {Alishov}, {Arbouch}, {Baransky}, {Barynova}, {Bai}, {Basa}, {Beradze},
  {Bertin}, {Berthier}, {Bla{\v{z}}ek}, {Bo{\"e}r}, {Burkhonov}, {Burrell},
  {Cailleau}, {Chabert}, {Chen}, {Christensen}, {Coleiro}, {Cordier}, {Corre},
  {Coughlin}, {Coward}, {Crisp}, {Delattre}, {Dietrich}, {Ducoin}, {Duverne},
  {Marchal-Duval}, {Gendre}, {Eymar}, {Fock-Hang}, {Han}, {Hello}, {Howell},
  {Inasaridze}, {Ismailov}, {Kann}, {Kapanadze}, {Klotz}, {Kochiashvili},
  {Lachaud}, {Leroy}, {Le Van Su}, {Lin}, {Li}, {Lognone}, {Marron}, {Mo},
  {Moore}, {Natsvlishvili}, {Noysena}, {Perrigault}, {Peyrot}, {Samadov},
  {Sadibekova}, {Simon}, {Stachie}, {Teng}, {Thierry}, {Th{\"o}ne}, {Tillayev},
  {Turpin}, {de Ugarte Postigo}, {Vachier}, {Vardosanidze}, {Vasylenko},
  {Vidadi}, {Wang}, {Wang}, {Wei}, {Yan}, {Zhang}, {Zhang}, \&
  {Zhang}}]{GRANDMAO3A}
{Antier}, S., {Agayeva}, S., {Aivazyan}, V., {et~al.} 2020{\natexlab{a}},
  \mnras, 492, 3904

\bibitem[{{Antier} {et~al.}(2020{\natexlab{b}}){Antier}, {Agayeva}, {Almualla},
  {Awiphan}, {Baransky}, {Barynova}, {Beradze}, {Bla{\v{z}}ek}, {Bo{\"e}r},
  {Burkhonov}, {Christensen}, {Coleiro}, {Corre}, {Coughlin}, {Crisp},
  {Dietrich}, {Ducoin}, {Duverne}, {Marchal-Duval}, {Gendre}, {Gokuldass},
  {Eggenstein}, {Eymar}, {Hello}, {Howell}, {Ismailov}, {Kann}, {Karpov},
  {Klotz}, {Kochiashvili}, {Lachaud}, {Leroy}, {Lin}, {Li}, {Ma{\v{s}}ek},
  {Mo}, {Menard}, {Morris}, {Noysena}, {Orange}, {Prouza}, {Rattanamala},
  {Sadibekova}, {Saint-Gelais}, {Serrau}, {Simon}, {Stachie}, {Th{\"o}ne},
  {Tillayev}, {Turpin}, {Postigo}, {Vasylenko}, {Vidadi}, {Was}, {Wang},
  {Zhang}, {Zhang}, \& {Zhang}}]{GRANDMA03B}
{Antier}, S., {Agayeva}, S., {Almualla}, M., {et~al.} 2020{\natexlab{b}},
  \mnras, 497, 5518

\bibitem[{{Antier} {et~al.}(2020{\natexlab{c}}){Antier}, {Agayeva}, {Almualla},
  {Awiphan}, {Baransky}, {Barynova}, {Beradze}, {Bla{\v{z}}ek}, {Bo{\"e}r},
  {Burkhonov}, {Christensen}, {Coleiro}, {Corre}, {Coughlin}, {Crisp},
  {Dietrich}, {Ducoin}, {Duverne}, {Marchal-Duval}, {Gendre}, {Gokuldass},
  {Eggenstein}, {Eymar}, {Hello}, {Howell}, {Ismailov}, {Kann}, {Karpov},
  {Klotz}, {Kochiashvili}, {Lachaud}, {Leroy}, {Lin}, {Li}, {Ma{\v{s}}ek},
  {Mo}, {Menard}, {Morris}, {Noysena}, {Orange}, {Prouza}, {Rattanamala},
  {Sadibekova}, {Saint-Gelais}, {Serrau}, {Simon}, {Stachie}, {Th{\"o}ne},
  {Tillayev}, {Turpin}, {Postigo}, {Vasylenko}, {Vidadi}, {Was}, {Wang},
  {Zhang}, {Zhang}, \& {Zhang}}]{2020MNRAS.497.5518A}
{Antier}, S., {Agayeva}, S., {Almualla}, M., {et~al.} 2020{\natexlab{c}},
  \mnras, 497, 5518

\bibitem[{{Arcavi} {et~al.}(2017){Arcavi}, {Howell}, {Kasen}, {Bildsten},
  {Hosseinzadeh}, {McCully}, {Wong}, {Katz}, {Gal-Yam}, {Sollerman}, {Taddia},
  {Leloudas}, {Fremling}, {Nugent}, {Horesh}, {Mooley}, {Rumsey}, {Cenko},
  {Graham}, {Perley}, {Nakar}, {Shaviv}, {Bromberg}, {Shen}, {Ofek}, {Cao},
  {Wang}, {Huang}, {Rui}, {Zhang}, {Li}, {Li}, {Zhang}, {Valenti}, {Guevel},
  {Shappee}, {Kochanek}, {Holoien}, {Filippenko}, {Fender}, {Nyholm}, {Yaron},
  {Kasliwal}, {Sullivan}, {Blagorodnova}, {Walters}, {Lunnan}, {Khazov},
  {Andreoni}, {Laher}, {Konidaris}, {Wozniak}, \& {Bue}}]{2017Natur.551..210A}
{Arcavi}, I., {Howell}, D.~A., {Kasen}, D., {et~al.} 2017, \nat, 551, 210

\bibitem[{{Aristidi} {et~al.}(2019){Aristidi}, {Ziad}, {Fantéï-Caujolle},
  {Renaud}, \& {Giordano}}]{AristidiEtAl2019a}
{Aristidi}, E., {Ziad}, A., {Fantéï-Caujolle}, Y., {Renaud}, C., \&
  {Giordano}, C. 2019, MNRAS, 486, 915

\bibitem[{{Arnouts} \& {Ilbert}(2011)}]{Arnouts2011}
{Arnouts}, S. \& {Ilbert}, O. 2011, {LePHARE: Photometric Analysis for Redshift
  Estimate}, Astrophysics Source Code Library, record ascl:1108.009

\bibitem[{{Ascenzi} {et~al.}(2020){Ascenzi}, {Oganesyan}, {Salafia},
  {Branchesi}, {Ghirlanda}, \& {Dall'Osso}}]{Ascenzi2020}
{Ascenzi}, S., {Oganesyan}, G., {Salafia}, O.~S., {et~al.} 2020, \aap, 641, A61

\bibitem[{Azzam {et~al.}(2020)Azzam, Ali, Elnagahy, Zead, Ahmed, Ismail, Saad,
  Shokry, Takey, Hendy, Mack, Yoshida, Kawabata, Akitaya, Darwish, Fouad,
  Helmy, Ismail, Elsayed, Ali, Abdel-sabour, Molham, Haroon, Osman, Hamdy,
  Issa, Abo-Elala, \& Essam}]{Azzam2020}
Azzam, Y.~A., Ali, G.~B., Elnagahy, F. I.~Y., {et~al.} 2020, in Ground-based
  and Airborne Instrumentation for Astronomy {VIII}, ed. C.~J. Evans, J.~J.
  Bryant, \& K.~Motohara ({SPIE})

\bibitem[{Azzam {et~al.}(2010)Azzam, Ali, Ismail, Haroon, \& Selim}]{azzam2010}
Azzam, Y.~A., Ali, G.~B., Ismail, H.~A., Haroon, A., \& Selim, I. 2010, in
  Proceedings of the Third UN/ESA/NASA Workshop on the International
  Heliophysical Year 2007 and Basic Space Science, ed. H.~J. Haubold \&
  A.~Mathai (Berlin, Heidelberg: Springer Berlin Heidelberg), 175--187

\bibitem[{{Bauswein} {et~al.}(2017){Bauswein}, {Just}, {Janka}, \&
  {Stergioulas}}]{2017ApJ...850L..34B}
{Bauswein}, A., {Just}, O., {Janka}, H.-T., \& {Stergioulas}, N. 2017, \apjl,
  850, L34

\bibitem[{{Belkin} {et~al.}(2022){Belkin}, {Pozanenko}, {Klunko}, {Pankov}, \&
  {GRB IKI FuN}}]{2022GCN.31773....1B}
{Belkin}, S., {Pozanenko}, A., {Klunko}, E., {Pankov}, N., \& {GRB IKI FuN}.
  2022, GRB Coordinates Network, 31773, 1

\bibitem[{{Bellm}(2014)}]{2014htu..conf...27B}
{Bellm}, E. 2014, in The Third Hot-wiring the Transient Universe Workshop, ed.
  P.~R. {Wozniak}, M.~J. {Graham}, A.~A. {Mahabal}, \& R.~{Seaman}, 27--33

\bibitem[{{Beniamini} {et~al.}(2020){Beniamini}, {Duque}, {Daigne}, \&
  {Mochkovitch}}]{Beniamini2020}
{Beniamini}, P., {Duque}, R., {Daigne}, F., \& {Mochkovitch}, R. 2020, \mnras,
  492, 2847

\bibitem[{{Benkhaldoun}(2018)}]{Benkhaldoun2018}
{Benkhaldoun}, Z. 2018, Nature Astronomy, 2, 352

\bibitem[{{Benkhaldoun} {et~al.}(2005){Benkhaldoun}, {Abahamid}, {El Azhari},
  \& {Lazrek}}]{Benkhaldoun2005}
{Benkhaldoun}, Z., {Abahamid}, A., {El Azhari}, Y., \& {Lazrek}, M. 2005, \aap,
  441, 839

\bibitem[{{Beradze} {et~al.}(2022{\natexlab{a}}){Beradze}, {Bhardwaj},
  {Culino}, {Hello}, {Masek}, {Raaijmakers}, {Rajabov}, {Sadibekova}, {Guo},
  {Wang}, {Zhu}, {Zhang}, {Kann}, {Noysena}, {Rinner}, {Benkhaldoun}, {Zhu},
  {Song}, {Antier}, {Duverne}, {Simon}, {Baransky}, {Godunova}, \& {Grandma
  Collaboration}}]{2022GCN.31884....1B}
{Beradze}, S., {Bhardwaj}, U., {Culino}, T., {et~al.} 2022{\natexlab{a}}, GRB
  Coordinates Network, 31884, 1

\bibitem[{{Beradze} {et~al.}(2022{\natexlab{b}}){Beradze}, {Bhardwaj},
  {Culino}, {Hello}, {Maske}, {Raaijmakers}, {Rajabov}, {Sadibekova}, {Guo},
  {Wang}, {Zhu}, {Zhang}, {Kann}, {Noysena}, {Kaouech}, {Rinner},
  {Benkhaldoun}, {Antier}, {Duverne}, {Freeberg}, {Hainich}, {Runger},
  {Karpov}, {Simon}, {Baransky}, {Godunova}, \& {Grandma
  Collaboration}}]{2022GCN.31903....1B}
{Beradze}, S., {Bhardwaj}, U., {Culino}, T., {et~al.} 2022{\natexlab{b}}, GRB
  Coordinates Network, 31903, 1

\bibitem[{{Bissaldi} {et~al.}(2022){Bissaldi}, {Meegan}, \& {Fermi GBM
  Team}}]{2022GCN.32054....1B}
{Bissaldi}, E., {Meegan}, C., \& {Fermi GBM Team}. 2022, GRB Coordinates
  Network, 32054, 1

\bibitem[{{Bizouard} {et~al.}(2022){Bizouard}, {Duverne}, {Iskandar},
  {Datashvili}, {Blazek}, {Turpin}, {Midavaine}, {Antier}, {Tosta Emelo},
  {Wang}, {Zhu}, {Song}, {Karpov}, {Kann}, {Marchais}, {Popowicz}, {Oksanen},
  {Serrau}, {Freeberg}, {Klotz}, {Kneip}, {Broens}, {Aguerre}, \& {Grandma
  Collaboration}}]{2022GCN.31785....1B}
{Bizouard}, M., {Duverne}, P.~A., {Iskandar}, A., {et~al.} 2022, GRB
  Coordinates Network, 31785, 1

\bibitem[{{Boquien} {et~al.}(2019){Boquien}, {Burgarella}, {Roehlly}, {Buat},
  {Ciesla}, {Corre}, {Inoue}, \& {Salas}}]{Boquien2019}
{Boquien}, M., {Burgarella}, D., {Roehlly}, Y., {et~al.} 2019, \aap, 622, A103

\bibitem[{{Coughlin} {et~al.}(2020{\natexlab{a}}){Coughlin}, {Antier},
  {Dietrich}, {Foley}, {Heinzel}, {Bulla}, {Christensen}, {Coulter}, {Issa}, \&
  {Khetan}}]{2020NatCo..11.4129C}
{Coughlin}, M.~W., {Antier}, S., {Dietrich}, T., {et~al.} 2020{\natexlab{a}},
  Nature Communications, 11, 4129

\bibitem[{Coughlin {et~al.}(2023)Coughlin, Bloom, Nir, Antier, du~Laz, van~der
  Walt, Crellin-Quick, Culino, Duev, Goldstein, Healy, Karambelkar, Lilleboe,
  Shin, Singer, Ahumada, Anand, Bellm, Dekany, Graham, Kasliwal, Kostadinova,
  Kiendrebeogo, Kulkarni, Jenkins, LeBaron, Neill, Parazin, Peloton, Riddle,
  Rusholme, van Santen, Sollerman, Stein, Turpin, Wold, Amat, Bonnefon,
  Bonnefoy, Flament, Kerkow, Kishore, Jani, Mahanty, Liu, Llinares, Makarison,
  Olliéric, Perez, Pont, \& Sharma}]{CoBl2023}
Coughlin, M.~W., Bloom, J.~S., Nir, G., {et~al.} 2023, A data science platform
  to enable time-domain astronomy

\bibitem[{{Coughlin} {et~al.}(2020{\natexlab{b}}){Coughlin}, {Dietrich},
  {Antier}, {Bulla}, {Foucart}, {Hotokezaka}, {Raaijmakers}, {Hinderer}, \&
  {Nissanke}}]{Coughlin2019}
{Coughlin}, M.~W., {Dietrich}, T., {Antier}, S., {et~al.} 2020{\natexlab{b}},
  \mnras, 492, 863

\bibitem[{{Coughlin} {et~al.}(2018){Coughlin}, {Dietrich}, Doctor, Kasen,
  Coughlin, Jerkstrand, Leloudas, McBrien, Metzger, O’Shaughnessy, \&
  Smartt}]{CoDi2018}
{Coughlin}, M.~W., {Dietrich}, T., Doctor, Z., {et~al.} 2018, Monthly Notices
  of the Royal Astronomical Society, 480, 3871

\bibitem[{Coughlin {et~al.}(2020)Coughlin, Dietrich, Heinzel, Khetan, Antier,
  Bulla, Christensen, Coulter, \& Foley}]{PhysRevResearch.2.022006}
Coughlin, M.~W., Dietrich, T., Heinzel, J., {et~al.} 2020, Phys. Rev. Research,
  2, 022006

\bibitem[{Coughlin {et~al.}(2019)Coughlin, Dietrich, Margalit, \&
  Metzger}]{CoDi2018b}
Coughlin, M.~W., Dietrich, T., Margalit, B., \& Metzger, B.~D. 2019, Monthly
  Notices of the Royal Astronomical Society: Letters, 489, L91

\bibitem[{{D'Ai} {et~al.}(2022){D'Ai}, {Ambrosi}, {D'Elia}, {Gronwall}, {Lien},
  {Page}, {Palmer}, \& {Neil Gehrels Swift Observatory
  Team}}]{2022GCN.31982....1D}
{D'Ai}, A., {Ambrosi}, E., {D'Elia}, V., {et~al.} 2022, GRB Coordinates
  Network, 31982, 1

\bibitem[{{de Ugarte Postigo} \& {Clavero Jimenez}(2022)}]{2022GCN.31775....1A}
{de Ugarte Postigo}, A. \& {Clavero Jimenez}, R. 2022, GRB Coordinates Network,
  31775, 1

\bibitem[{{de Wet} {et~al.}(2022){de Wet}, {Vreeswijk}, {Malesani}, \&
  {Meerlicht Consortium}}]{2022GCN.32044....1D}
{de Wet}, S., {Vreeswijk}, P.~M., {Malesani}, D.~B., \& {Meerlicht Consortium}.
  2022, GRB Coordinates Network, 32044, 1

\bibitem[{{Dornic} {et~al.}(2022){Dornic}, {Klotz}, {Sabahaddin}, {Midavaine},
  {Rupchandani}, {Duverne}, {Rajabov}, {Song}, {Wang}, {Zhu}, {Wang}, {Zeng},
  {Iskandar}, {Thierry}, {Kaeouach}, {Benkhaldoun}, {Antier}, {de Ugarte
  Postigo}, \& {Grandma Collaboration}}]{2022GCN.31977....1D}
{Dornic}, D., {Klotz}, A., {Sabahaddin}, A., {et~al.} 2022, GRB Coordinates
  Network, 31977, 1

\bibitem[{Duverne {et~al.}(2022)Duverne, Antier, Basa, Corre, Coughlin,
  Filippenko, Klotz, Hello, \& Zheng}]{muphoten}
Duverne, P.~A., Antier, S., Basa, S., {et~al.} 2022, Publications of the
  Astronomical Society of the Pacific, 134, 114504

\bibitem[{{Ferro} {et~al.}(2022){Ferro}, {Bernardini}, {Brivio}, {D'Avanzo},
  {Gropp}, {Kennea}, {Kuin}, {Laha}, {Marshall}, {Page}, {Palmer}, {Parsotan},
  {Sbarrato}, {Sbarufatti}, {Siegel}, \& {Neil Gehrels Swift Observatory
  Team}}]{2022GCN.31787....1F}
{Ferro}, M., {Bernardini}, M.~G., {Brivio}, R., {et~al.} 2022, GRB Coordinates
  Network, 31787, 1

\bibitem[{{Genet} {et~al.}(2007){Genet}, {Daigne}, \& {Mochkovitch}}]{genet:07}
{Genet}, F., {Daigne}, F., \& {Mochkovitch}, R. 2007, \mnras, 381, 732

\bibitem[{{Gokuldass} {et~al.}(2021){Gokuldass}, {Morris}, {Orange},
  {Cucchiara}, \& {Strausbaugh}}]{2021AAS...23713501G}
{Gokuldass}, P., {Morris}, D., {Orange}, N., {Cucchiara}, A., \& {Strausbaugh},
  R. 2021, in American Astronomical Society Meeting Abstracts, Vol.~53,
  American Astronomical Society Meeting Abstracts, 135.01

\bibitem[{Goldstein {et~al.}(2017)Goldstein, Veres, Burns, Briggs, Hamburg,
  Kocevski, Wilson-Hodge, Preece, Poolakkil, Roberts, Hui, Connaughton,
  Racusin, Kienlin, Canton, Christensen, Littenberg, Siellez, Blackburn,
  Broida, Bissaldi, Cleveland, Gibby, Giles, Kippen, McBreen, McEnery, Meegan,
  Paciesas, \& Stanbro}]{goldstein_ordinary_2017}
Goldstein, A., Veres, P., Burns, E., {et~al.} 2017, The Astrophysical Journal,
  848, L14

\bibitem[{{Gopalakrishnan} {et~al.}(2022){Gopalakrishnan}, {Prasad},
  {Waratkar}, {Vibhute}, {Bhalerao}, {Bhattacharya}, {Rao}, {Vadawale}, \&
  {AstroSat CZTI Collaboration}}]{2022GCN.32070....1G}
{Gopalakrishnan}, R., {Prasad}, V., {Waratkar}, G., {et~al.} 2022, GRB
  Coordinates Network, 32070, 1

\bibitem[{Granot \& Kumar(2006)}]{Granot2006}
Granot, J. \& Kumar, P. 2006, Monthly Notices of the Royal Astronomical
  Society: Letters, 366, L13

\bibitem[{{Granot} \& {Sari}(2002)}]{2002ApJ...568..820G}
{Granot}, J. \& {Sari}, R. 2002, \apj, 568, 820

\bibitem[{{Greiner} {et~al.}(2008){Greiner}, {Bornemann}, {Clemens}, {Deuter},
  {Hasinger}, {Honsberg}, {Huber}, {Huber}, {Krauss}, {Kr{\"u}hler},
  {K{\"u}pc{\"u} Yolda{\c{s}}}, {Mayer-Hasselwander}, {Mican}, {Primak},
  {Schrey}, {Steiner}, {Szokoly}, {Th{\"o}ne}, {Yolda{\c{s}}}, {Klose}, {Laux},
  \& {Winkler}}]{Greiner2008}
{Greiner}, J., {Bornemann}, W., {Clemens}, C., {et~al.} 2008, \pasp, 120, 405

\bibitem[{{Groot} {et~al.}(2022){Groot}, {de Wet}, {Malesani}, {Levan},
  {Vreeswijk}, \& {Meerlicht Consortium}}]{2022GCN.31974....1G}
{Groot}, P.~J., {de Wet}, S., {Malesani}, D.~B., {et~al.} 2022, GRB Coordinates
  Network, 31974, 1

\bibitem[{{Gupta} {et~al.}(2022){Gupta}, {Ror}, {Kumar}, {Dimple}, {Ghosh},
  {Aryan}, {Chand}, {Pandey}, \& {Misra}}]{2022GCN.31793....1G}
{Gupta}, R., {Ror}, A., {Kumar}, A., {et~al.} 2022, GRB Coordinates Network,
  31793, 1

\bibitem[{{Haggard} {et~al.}(2017){Haggard}, {Nynka}, {Ruan}, {Kalogera},
  {Cenko}, {Evans}, \& {Kennea}}]{2017ApJ...848L..25H}
{Haggard}, D., {Nynka}, M., {Ruan}, J.~J., {et~al.} 2017, \apjl, 848, L25

\bibitem[{{Hallinan} {et~al.}(2017){Hallinan}, {Corsi}, {Mooley}, {Hotokezaka},
  {Nakar}, {Kasliwal}, {Kaplan}, {Frail}, {Myers}, {Murphy}, {De}, {Dobie},
  {Allison}, {Bannister}, {Bhalerao}, {Chandra}, {Clarke}, {Giacintucci}, {Ho},
  {Horesh}, {Kassim}, {Kulkarni}, {Lenc}, {Lockman}, {Lynch}, {Nichols},
  {Nissanke}, {Palliyaguru}, {Peters}, {Piran}, {Rana}, {Sadler}, \&
  {Singer}}]{Hallinan_2017Sci}
{Hallinan}, G., {Corsi}, A., {Mooley}, K.~P., {et~al.} 2017, Science, 358, 1579

\bibitem[{{Hosokawa} {et~al.}(2022){Hosokawa}, {Imai}, {Murata}, {Niwano},
  {Ito}, {Takamatsu}, {Noto}, {Sato}, {Takaku}, {Yamaguchi}, {Yatsu}, {Kawai},
  \& {MITSuME Collaboration}}]{2022GCN.31776...1H}
{Hosokawa}, R., {Imai}, Y., {Murata}, K.~L., {et~al.} 2022, GRB Coordinates
  Network, 31776, 1

\bibitem[{{Hotokezaka} {et~al.}(2019){Hotokezaka}, {Nakar}, {Gottlieb},
  {Nissanke}, {Masuda}, {Hallinan}, {Mooley}, \&
  {Deller}}]{2019NatAs...3..940H}
{Hotokezaka}, K., {Nakar}, E., {Gottlieb}, O., {et~al.} 2019, Nature Astronomy,
  3, 940

\bibitem[{{Hu} {et~al.}(2017){Hu}, {Wu}, {Andreoni}, {Ashley}, {Cooke}, {Cui},
  {Du}, {Dai}, {Gu}, {Hu}, {Lu}, {Li}, {Li}, {Liang}, {Liu}, {Ma}, {Shang},
  {Sun}, {Suntzeff}, {Tao}, {Udden}, {Wang}, {Wang}, {Wen}, {Xiao}, {Su},
  {Yang}, {Yang}, {Yuan}, {Zhou}, {Zhang}, {Zhou}, \&
  {Zhu}}]{2017SciBu..62.1433H}
{Hu}, L., {Wu}, X., {Andreoni}, I., {et~al.} 2017, Science Bulletin, 62, 1433

\bibitem[{{Hu} {et~al.}(2022{\natexlab{a}}){Hu}, {Fernandez-Garcia}, {Sun},
  {Castro-Tirado}, {Caballero-Garcia}, {Sanchez-Ramirez}, {Perez-Garcia},
  {Perez del Pulgar}, {Castellon}, {Carrasco}, {Reina}, \&
  {Rendon}}]{2022GCN.31990....1H}
{Hu}, Y.~D., {Fernandez-Garcia}, E., {Sun}, T.~R., {et~al.} 2022{\natexlab{a}},
  GRB Coordinates Network, 31990, 1

\bibitem[{{Hu} {et~al.}(2022{\natexlab{b}}){Hu}, {Fernandez-Garcia}, {Sun},
  {Castro-Tirado}, {Sanchez-Ramirez}, {Caballero-Garcia}, {Hiriart}, {Lee},
  {Perez del Pulgar}, {Carrasco}, \& {Park}}]{2022GCN.31859....1H}
{Hu}, Y.~D., {Fernandez-Garcia}, E., {Sun}, T.~R., {et~al.} 2022{\natexlab{b}},
  GRB Coordinates Network, 31859, 1

\bibitem[{{Hu} {et~al.}(2022{\natexlab{c}}){Hu}, {Sun}, {Caballero-Garcia},
  {Sanchez-Ramirez}, {Perez-Garcia}, {Castro-Tirado}, {Minguez}, \&
  {Vico}}]{2022GCN.31980....1H}
{Hu}, Y.~D., {Sun}, T.~R., {Caballero-Garcia}, M.~D., {et~al.}
  2022{\natexlab{c}}, GRB Coordinates Network, 31980, 1

\bibitem[{{Hunt} {et~al.}(2014){Hunt}, {Palazzi}, {Micha{\l}owski}, {Rossi},
  {Savaglio}, {Basa}, {Berta}, {Bianchi}, {Covino}, {D'Elia}, {Ferrero},
  {G{\"o}tz}, {Greiner}, {Klose}, {Le Borgne}, {Le Floc'h}, {Pian},
  {Piranomonte}, {Schady}, \& {Vergani}}]{2014A&A...565A.112H}
{Hunt}, L.~K., {Palazzi}, E., {Micha{\l}owski}, M.~J., {et~al.} 2014, \aap,
  565, A112

\bibitem[{Im {et~al.}(2010)Im, Ko, Cho, Choi, Jeon, Lee, \&
  Ibrahimov}]{Im2010SEOULNU}
Im, M., Ko, J., Cho, Y., {et~al.} 2010, Journal of the Korean Astronomical
  Society, 43, 75

\bibitem[{{Jiang} {et~al.}(2022{\natexlab{a}}){Jiang}, {Zhu}, {Fu}, {Liu},
  {Xu}, {Gao}, \& {Liu}}]{2022GCN.31998....1J}
{Jiang}, S.~Q., {Zhu}, Z.~P., {Fu}, S.~Y., {et~al.} 2022{\natexlab{a}}, GRB
  Coordinates Network, 31998, 1

\bibitem[{{Jiang} {et~al.}(2022{\natexlab{b}}){Jiang}, {Zhu}, {Fu}, {Liu},
  {Xu}, {Gao}, \& {Liu}}]{2022GCN.31999....1J}
{Jiang}, S.~Q., {Zhu}, Z.~P., {Fu}, S.~Y., {et~al.} 2022{\natexlab{b}}, GRB
  Coordinates Network, 31999, 1

\bibitem[{{Kann} {et~al.}(2023){Kann}, {Agayeva}, {Aivazyan}, {Alishov},
  {Andrade}, {Antier}, {Baransky}, {Bendjoya}, {Benkhaldoun}, {Beradze},
  {Berezin}, {Bo{\"e}r}, {Broens}, {Brunier}, {Bulla}, {Burkhonov}, {Burns},
  {Chen}, {Chen}, {Conti}, {Coughlin}, {Cui}, {Daigne}, {Delaveau},
  {Devillepoix}, {Dietrich}, {Dornic}, {Dubois}, {Ducoin}, {Durand}, {Duverne},
  {Eggenstein}, {Ehgamberdiev}, {Fouad}, {Freeberg}, {Froebrich}, {Ge},
  {Gervasoni}, {Godunova}, {Gokuldass}, {Gurbanov}, {Han}, {Hasanov}, {Hello},
  {Hussenot-Desenonges}, {Inasaridze}, {Iskandar}, {Ismailov}, {Janati}, {Jegou
  du Laz}, {Jia}, {Karpov}, {Kaeouach}, {Kiendrebeogo}, {Klotz}, {Kneip},
  {Kochiashvili}, {Kunert}, {Lekic}, {Leonini}, {Li}, {Li}, {Li}, {Liao},
  {Logie}, {Lu}, {Mao}, {Marchais}, {M{\'e}nard}, {Morris}, {Natsvlishvili},
  {Nedora}, {Noonan}, {Noysena}, {Orange}, {Pang}, {Peng}, {Pellouin},
  {Peloton}, {Pradier}, {Pyshna}, {Rajabo}, {Rau}, {Rinner}, {Rivet},
  {Romanov}, {Rosi}, {Rupchandani}, {Serrau}, {Shokry}, {Simon}, {Smith},
  {Sokoliuk}, {Soliman}, {Song}, {Takey}, {Tillayev}, {Tinjaca Ramirez}, {Tosta
  e Melo}, {Turpin}, {de Ugarte Postigo}, {Vanaverbeke}, {Vasylenko}, {Vernet},
  {Vidadi}, {Wang}, {Wang}, {Wang}, {Wang}, {Xiong}, {Xu}, {Xue}, {Zeng},
  {Zhang}, {Zhao}, \& {Zhao}}]{2023arXiv230206225K}
{Kann}, D.~A., {Agayeva}, S., {Aivazyan}, V., {et~al.} 2023, arXiv e-prints,
  arXiv:2302.06225

\bibitem[{{Karpov}(2021)}]{stdpipe}
{Karpov}, S. 2021, {STDPipe: Simple Transient Detection Pipeline}

\bibitem[{{Klingler} {et~al.}(2022{\natexlab{a}}){Klingler}, {Ambrosi},
  {D'Elia}, {Dichiara}, {Gropp}, {Krimm}, {Marshall}, {Palmer}, {Parsotan},
  {Sakamoto}, \& {Neil Gehrels Swift Observatory Team}}]{2022GCN.31820....1K}
{Klingler}, N.~J., {Ambrosi}, E., {D'Elia}, V., {et~al.} 2022{\natexlab{a}},
  GRB Coordinates Network, 31820, 1

\bibitem[{{Klingler} {et~al.}(2022{\natexlab{b}}){Klingler}, {Gropp}, {Page},
  {Palmer}, {Parsotan}, {Sbarufatti}, {Tohuvavohu}, \& {Neil Gehrels Swift
  Observatory Team}}]{2022GCN.31881....1K}
{Klingler}, N.~J., {Gropp}, J.~D., {Page}, K.~L., {et~al.} 2022{\natexlab{b}},
  GRB Coordinates Network, 31881, 1

\bibitem[{{Knust} {et~al.}(2017){Knust}, {Greiner}, {van Eerten}, {Schady},
  {Kann}, {Chen}, {Delvaux}, {Graham}, {Klose}, {Kr{\"u}hler}, {McConnell},
  {Nicuesa Guelbenzu}, {Perley}, {Schmidl}, {Schweyer}, {Tanga}, \&
  {Varela}}]{2017A&A...607A..84K}
{Knust}, F., {Greiner}, J., {van Eerten}, H.~J., {et~al.} 2017, \aap, 607, A84

\bibitem[{{Kuin} {et~al.}(2022){Kuin}, {Ferro}, \& {Swift/UVOT
  Team}}]{2022GCN.31801....1K}
{Kuin}, N.~P.~M., {Ferro}, M., \& {Swift/UVOT Team}. 2022, GRB Coordinates
  Network, 31801, 1

\bibitem[{{Kumar} {et~al.}(2008{\natexlab{a}}){Kumar}, {Narayan}, \&
  {Johnson}}]{2008MNRAS.388.1729K}
{Kumar}, P., {Narayan}, R., \& {Johnson}, J.~L. 2008{\natexlab{a}}, \mnras,
  388, 1729

\bibitem[{{Kumar} {et~al.}(2008{\natexlab{b}}){Kumar}, {Narayan}, \&
  {Johnson}}]{2008Sci...321..376K}
{Kumar}, P., {Narayan}, R., \& {Johnson}, J.~L. 2008{\natexlab{b}}, Science,
  321, 376

\bibitem[{{Lien} {et~al.}(2022){Lien}, {Barthelmy}, {Klingler}, {Krimm},
  {Laha}, {Lien}, {Markwardt}, {Palmer}, {Parsotan}, {Sakamoto}, \&
  {Stamatikos}}]{2022GCN.31834....1L}
{Lien}, A.~Y., {Barthelmy}, S.~D., {Klingler}, N.~J., {et~al.} 2022, GRB
  Coordinates Network, 31834, 1

\bibitem[{{Lipunov} {et~al.}(2022{\natexlab{a}}){Lipunov}, {Kornilov},
  {Gorbovskoy}, {Zhirkov}, {Tyurina}, {Balanutsa}, {Kuznetsov}, {Vlasenko},
  {Antipov}, {Zimnukhov}, {Senik}, {Minkina}, {Chasovnikov}, {Topolev},
  {Kuvshinov}, {Cheryasov}, {Kechin}, {Podesta}, {Lopez}, {Podesta},
  {Francile}, {Rebolo}, {Serra}, {Buckley}, {Gres}, {Budnev}, {Carrasco},
  {Valdes}, {Chavushyan}, {Patino Alvarez}, {Martinez}, {Corella}, {Rodriguez},
  {Tlatov}, {Dormidontov}, {Gabovich}, \& {Yurkov}}]{2022GCN.32042....1L}
{Lipunov}, V., {Kornilov}, V., {Gorbovskoy}, E., {et~al.} 2022{\natexlab{a}},
  GRB Coordinates Network, 32042, 1

\bibitem[{{Lipunov} {et~al.}(2022{\natexlab{b}}){Lipunov}, {Kuznetsov},
  {Gress}, {Gorbovskoy}, {Tyurina}, {Balanutsa}, {Zhirkov}, {Chasovnikov},
  {Vlasenko}, {Antipov}, {Senik}, {Kuvshinov}, {Topolev}, {Kechin}, {Tselik},
  {Francile}, {Podesta}, {Lopez}, {Podesta}, {Buckley}, {Rebolo}, {Serra},
  {Carrasco}, {Valdes}, {Chavushyan}, {Patino Alvarez}, {Martinez}, {Corella},
  {Rodriguez}, {Budnev}, {Tlatov}, {Dormidontov}, {Gabovich}, \&
  {Yurkov}}]{2022GCN.31770....1L}
{Lipunov}, V., {Kuznetsov}, A., {Gress}, O., {et~al.} 2022{\natexlab{b}}, GRB
  Coordinates Network, 31770, 1

\bibitem[{{Lipunov} {et~al.}(2022{\natexlab{c}}){Lipunov}, {Vlasenko},
  {Kuznetsov}, {Gress}, {Gorbovskoy}, {Tyurina}, {Balanutsa}, {Zhirkov},
  {Chasovnikov}, {Antipov}, {Senik}, {Kuvshinov}, {Topolev}, {Kechin},
  {Tselik}, {Francile}, {Podesta}, {Lopez}, {Podesta}, {Buckley}, {Rebolo},
  {Serra}, {Carrasco}, {Valdes}, {Chavushyan}, {Patino Alvarez}, {Martinez},
  {Corella}, {Rodriguez}, {Budnev}, {Tlatov}, {Dormidontov}, {Gabovich}, \&
  {Yurkov}}]{2022GCN.31789....1L}
{Lipunov}, V., {Vlasenko}, D., {Kuznetsov}, A., {et~al.} 2022{\natexlab{c}},
  GRB Coordinates Network, 31789, 1

\bibitem[{{Mao} {et~al.}(2012){Mao}, {Malesani}, {D'Avanzo}, {Covino}, {Li},
  {Jakobsson}, \& {Bai}}]{mao12}
{Mao}, J., {Malesani}, D., {D'Avanzo}, P., {et~al.} 2012, \aap, 538, A1

\bibitem[{{Margalit} \& {Metzger}(2017)}]{2017ApJ...850L..19M}
{Margalit}, B. \& {Metzger}, B.~D. 2017, \apjl, 850, L19

\bibitem[{{M{\"o}ller} {et~al.}(2021){M{\"o}ller}, {Peloton}, {Ishida},
  {Arnault}, {Bachelet}, {Blaineau}, {Boutigny}, {Chauhan}, {Gangler},
  {Hernandez}, {Hrivnac}, {Leoni}, {Leroy}, {Moniez}, {Pateyron}, {Ramparison},
  {Turpin}, {Ansari}, {Allam}, {Bajat}, {Biswas}, {Boucaud}, {Bregeon},
  {Campagne}, {Cohen-Tanugi}, {Coleiro}, {Dornic}, {Fouchez}, {Godet}, {Gris},
  {Karpov}, {Nebot Gomez-Moran}, {Neveu}, {Plaszczynski}, {Savchenko}, \&
  {Webb}}]{2021MNRAS.501.3272M}
{M{\"o}ller}, A., {Peloton}, J., {Ishida}, E. E.~O., {et~al.} 2021, \mnras,
  501, 3272

\bibitem[{{Murata} {et~al.}(2022){Murata}, {Hosokawa}, {Imai}, {Ito}, {Sasada},
  {Niwano}, {Takamatsu}, {Sato}, {Tateda}, {Hattori}, {Yatsu}, {Kawai}, \&
  {MITSuME Collaboration}}]{2022GCN.32059....1M}
{Murata}, K.~L., {Hosokawa}, R., {Imai}, Y., {et~al.} 2022, GRB Coordinates
  Network, 32059, 1

\bibitem[{{Neff} {et~al.}(2004){Neff}, {Allen}, {Aurin}, {Boyajian},
  {Crowther}, {Davis}, {Drost}, {Giblin}, {Hurley}, {Lucas}, {Nations},
  {Smith}, {Thomas}, \& {Walsh}}]{2004AN....325..669N}
{Neff}, J.~E., {Allen}, D.~K., {Aurin}, D.~M., {et~al.} 2004, Astronomische
  Nachrichten, 325, 669

\bibitem[{{Nicuesa Guelbenzu} {et~al.}(2022){Nicuesa Guelbenzu}, {Klose},
  {Stecklum}, {Melnikov}, \& {Laux}}]{2022GCN.31797....1N}
{Nicuesa Guelbenzu}, A., {Klose}, S., {Stecklum}, B., {Melnikov}, S., \&
  {Laux}, U. 2022, GRB Coordinates Network, 31797, 1

\bibitem[{{Oates} {et~al.}(2022){Oates}, {Caputo}, \& {Swift/UVOT
  Team}}]{2022GCN.31855....1O}
{Oates}, S.~R., {Caputo}, R., \& {Swift/UVOT Team}. 2022, GRB Coordinates
  Network, 31855, 1

\bibitem[{{Oganesyan} {et~al.}(2020){Oganesyan}, {Ascenzi}, {Branchesi},
  {Salafia}, {Dall'Osso}, \& {Ghirlanda}}]{Oganesyan2020}
{Oganesyan}, G., {Ascenzi}, S., {Branchesi}, M., {et~al.} 2020, \apj, 893, 88

\bibitem[{{Page} {et~al.}(2022){Page}, {Barthelmy}, {Gropp}, {Kennea}, {Lien},
  {Palmer}, {Parsotan}, \& {Neil Gehrels Swift Observatory
  Team}}]{2022GCN.31769....1P}
{Page}, K.~L., {Barthelmy}, S.~D., {Gropp}, J.~D., {et~al.} 2022, GRB
  Coordinates Network, 31769, 1

\bibitem[{{Palmerio} {et~al.}(2019){Palmerio}, {Vergani}, {Salvaterra},
  {Sanders}, {Japelj}, {Vidal-Garc{\'\i}a}, {D'Avanzo}, {Corre}, {Perley},
  {Shapley}, {Boissier}, {Greiner}, {Le Floc'h}, \&
  {Wiseman}}]{2019A&A...623A..26P}
{Palmerio}, J.~T., {Vergani}, S.~D., {Salvaterra}, R., {et~al.} 2019, \aap,
  623, A26

\bibitem[{{Pankov} {et~al.}(2022){Pankov}, {Belkin}, {Pozanenko}, {Klunko},
  {Inasaridze}, \& {GRB IKI FuN}}]{2022GCN.31976....1P}
{Pankov}, N., {Belkin}, S., {Pozanenko}, A., {et~al.} 2022, GRB Coordinates
  Network, 31976, 1

\bibitem[{{Perego} {et~al.}(2021){Perego}, {Thielemann}, \&
  {Cescutti}}]{2021hgwa.bookE..13P}
{Perego}, A., {Thielemann}, F.~K., \& {Cescutti}, G. 2021, in Handbook of
  Gravitational Wave Astronomy (Springer), 13

\bibitem[{{Perley} {et~al.}(2016){Perley}, {Niino}, {Tanvir}, {Vergani}, \&
  {Fynbo}}]{2016SSRv..202..111P}
{Perley}, D.~A., {Niino}, Y., {Tanvir}, N.~R., {Vergani}, S.~D., \& {Fynbo}, J.
  P.~U. 2016, \ssr, 202, 111

\bibitem[{Savchenko {et~al.}(2017)Savchenko, Ferrigno, Kuulkers, Bazzano,
  Bozzo, Brandt, Chenevez, Courvoisier, Diehl, Domingo, Hanlon, Jourdain, von
  Kienlin, Laurent, Lebrun, Lutovinov, Martin-Carrillo, Mereghetti, Natalucci,
  Rodi, Roques, Sunyaev, \& Ubertini}]{savchenko_integral_2017}
Savchenko, V., Ferrigno, C., Kuulkers, E., {et~al.} 2017, The Astrophysical
  Journal, 848, L15

\bibitem[{{Schlafly} \& {Finkbeiner}(2011)}]{Schlafly2011}
{Schlafly}, E.~F. \& {Finkbeiner}, D.~P. 2011, \apj, 737, 103

\bibitem[{{Schlegel} {et~al.}(1998){Schlegel}, {Finkbeiner}, \&
  {Davis}}]{Schlegel1998}
{Schlegel}, D.~J., {Finkbeiner}, D.~P., \& {Davis}, M. 1998, \apj, 500, 525

\bibitem[{{Schneider} {et~al.}(2022){Schneider}, {Le Floc'h}, {Arabsalmani},
  {Vergani}, \& {Palmerio}}]{2022A&A...666A..14S}
{Schneider}, B., {Le Floc'h}, E., {Arabsalmani}, M., {Vergani}, S.~D., \&
  {Palmerio}, J.~T. 2022, \aap, 666, A14

\bibitem[{Sergeev {et~al.}(2014)Sergeev, Burkhonov, Dudinov, Zheleznyak,
  Krugly, Molotov, Shulga, Ehgamberdiev, Konichek, \& Kochetov}]{Sergeev2014}
Sergeev, A.~V., Burkhonov, O.~A., Dudinov, V.~N., {et~al.} 2014, Radio Physics
  and Radio Astronomy, 19, 20

\bibitem[{{Song} {et~al.}(2022){Song}, {Zhu}, {Melo}, {Ducoin}, {Guessoum},
  {Corradi}, {Culino}, {Gurbanov}, {Hesenov}, {de Ugarte Postigo}, {Antier},
  {Wang}, {Iskandar}, {Wang}, {Zeng}, {Marchais}, {Duverne}, {Leroy}, {Kann},
  {Simon}, {Baransky}, {Godunova}, \& {Grandma
  Collaboration}}]{2022GCN.31883....1S}
{Song}, X., {Zhu}, J., {Melo}, I. T.~E., {et~al.} 2022, GRB Coordinates
  Network, 31883, 1

\bibitem[{{Strausbaugh} \&
  {Cucchiara}(2022{\natexlab{a}})}]{2022GCN.31774....1S}
{Strausbaugh}, R. \& {Cucchiara}, A. 2022{\natexlab{a}}, GRB Coordinates
  Network, 31774, 1

\bibitem[{{Strausbaugh} \&
  {Cucchiara}(2022{\natexlab{b}})}]{2022GCN.31795....1S}
{Strausbaugh}, R. \& {Cucchiara}, A. 2022{\natexlab{b}}, GRB Coordinates
  Network, 31795, 1

\bibitem[{{Strausbaugh} \&
  {Cucchiara}(2022{\natexlab{c}})}]{2022GCN.31862....1S}
{Strausbaugh}, R. \& {Cucchiara}, A. 2022{\natexlab{c}}, GRB Coordinates
  Network, 31862, 1

\bibitem[{{Swain} {et~al.}(2022){Swain}, {Kumar}, {Angail}, {Bhalerao},
  {Anupama}, {Barway}, \& {GIT Team}}]{2022GCN.31997....1S}
{Swain}, V., {Kumar}, H., {Angail}, K., {et~al.} 2022, GRB Coordinates Network,
  31997, 1

\bibitem[{{Tang} {et~al.}(2019){Tang}, {Huang}, {Geng}, \&
  {Zhang}}]{2019ApJS..245....1T}
{Tang}, C.-H., {Huang}, Y.-F., {Geng}, J.-J., \& {Zhang}, Z.-B. 2019, \apjs,
  245, 1

\bibitem[{{Uhm} \& {Beloborodov}(2007)}]{uhm:07}
{Uhm}, Z.~L. \& {Beloborodov}, A.~M. 2007, \apjl, 665, L93

\bibitem[{{Uhm} {et~al.}(2012){Uhm}, {Zhang}, {Hasco{\"e}t}, {Daigne},
  {Mochkovitch}, \& {Park}}]{uhm:12}
{Uhm}, Z.~L., {Zhang}, B., {Hasco{\"e}t}, R., {et~al.} 2012, \apj, 761, 147

\bibitem[{{Watson} {et~al.}(2022{\natexlab{a}}){Watson}, {Butler}, {Kutyrev},
  {Lee}, {Troja}, {Richer}, {Fox}, {Prochaska}, {Bloom}, {Ramirez-Ruiz},
  {Gonz{\'a}lez}, {Rom{\'a}n-Z{\'u}{\~n}iga}, {Moseley}, {Becerra}, {Dichiara},
  \& {L{\'o}pez}}]{2022GCN.31870....1W}
{Watson}, A.~M., {Butler}, N., {Kutyrev}, A., {et~al.} 2022{\natexlab{a}}, GRB
  Coordinates Network, 31870, 1

\bibitem[{{Watson} {et~al.}(2022{\natexlab{b}}){Watson}, {Butler}, {Troja},
  {Kutyrev}, {Lee}, {Richer}, {Fox}, {Prochaska}, {Bloom}, {Ramirez-Ruiz},
  {Gonz{\'a}lez}, {Rom{\'a}n-Z{\'u}{\~n}iga}, {Moseley}, {Becerra}, {Dichiara},
  \& {L{\'o}pez}}]{2022GCN.31902....1W}
{Watson}, A.~M., {Butler}, N., {Troja}, E., {et~al.} 2022{\natexlab{b}}, GRB
  Coordinates Network, 31902, 1

\bibitem[{{Watson} {et~al.}(2022{\natexlab{c}}){Watson}, {Butler}, {Troja},
  {Kutyrev}, {Lee}, {Richer}, {Fox}, {Prochaska}, {Bloom}, {Ramirez-Ruiz},
  {Gonz{\'a}lez}, {Rom{\'a}n-Z{\'u}{\~n}iga}, {Moseley}, {Becerra}, {Dichiara},
  \& {L{\'o}pez}}]{2022GCN.31898....1W}
{Watson}, A.~M., {Butler}, N., {Troja}, E., {et~al.} 2022{\natexlab{c}}, GRB
  Coordinates Network, 31898, 1

\bibitem[{{Watson} {et~al.}(2022{\natexlab{d}}){Watson}, {Troja}, {Butler},
  {Kutyrev}, {Lee}, {Richer}, {Fox}, {Prochaska}, {Bloom}, {Ramirez-Ruiz},
  {Gonz{\'a}lez}, {Rom{\'a}n-Z{\'u}{\~n}iga}, {Moseley}, {Becerra}, {Dichiara},
  \& {L{\'o}pez}}]{2022GCN.31882....1W}
{Watson}, A.~M., {Troja}, E., {Butler}, N., {et~al.} 2022{\natexlab{d}}, GRB
  Coordinates Network, 31882, 1

\bibitem[{{Wei} {et~al.}(2016){Wei}, {Cordier}, {Antier}, {Antilogus},
  {Atteia}, {Bajat}, {Basa}, {Beckmann}, {Bernardini}, {Boissier}, {Bouchet},
  {Burwitz}, {Claret}, {Dai}, {Daigne}, {Deng}, {Dornic}, {Feng}, {Foglizzo},
  {Gao}, {Gehrels}, {Godet}, {Goldwurm}, {Gonzalez}, {Gosset}, {G{\"o}tz},
  {Gouiffes}, {Grise}, {Gros}, {Guilet}, {Han}, {Huang}, {Huang}, {Jouret},
  {Klotz}, {La Marle}, {Lachaud}, {Le Floch}, {Lee}, {Leroy}, {Li}, {Li}, {Li},
  {Liang}, {Lyu}, {Mercier}, {Migliori}, {Mochkovitch}, {O'Brien}, {Osborne},
  {Paul}, {Perinati}, {Petitjean}, {Piron}, {Qiu}, {Rau}, {Rodriguez},
  {Schanne}, {Tanvir}, {Vangioni}, {Vergani}, {Wang}, {Wang}, {Wang}, {Wang},
  {Watson}, {Webb}, {Wei}, {Willingale}, {Wu}, {Wu}, {Xin}, {Xu}, {Yu}, {Yu},
  {Yu}, {Zhang}, {Zhang}, {Zhang}, \& {Zhou}}]{2016arXiv161006892W}
{Wei}, J., {Cordier}, B., {Antier}, S., {et~al.} 2016, arXiv e-prints,
  arXiv:1610.06892

\bibitem[{{Yan} {et~al.}(2022){Yan}, {Andrade}, {Beradze}, {Duverne},
  {Kruiswijk}, {Raaijmakers}, {Vardosanidze}, {Turpin}, {Pilloix}, {Antier},
  {de Ugarte Postigo}, {Kann}, {Simon}, {Baransky}, {Godunova}, {Inasaridze},
  {Natsvlishvili}, {Kochiashvili}, {Aivazyan}, {Kapanadze}, {Datashvili},
  {Rinner}, {Benkhaldoun}, {Freeberg}, {Kaouech}, {Guo}, {Wang}, {Wang},
  {Wang}, {Karpov}, {Masek}, {Prouza}, \& {Grandma
  Collaboration}}]{Yan2022GCN32058}
{Yan}, S., {Andrade}, C., {Beradze}, S., {et~al.} 2022, GRB Coordinates
  Network, 32058, 1

\bibitem[{{Zhao} {et~al.}(2020){Zhao}, {Liu}, {Gao}, {Lan}, {Lei}, \&
  {Xie}}]{2020ApJ...896...42Z}
{Zhao}, L., {Liu}, L., {Gao}, H., {et~al.} 2020, \apj, 896, 42

\bibitem[{{Zheng} {et~al.}(2022){Zheng}, {Filippenko}, \& {KAIT GRB
  Team}}]{2022GCN.32051....1Z}
{Zheng}, W., {Filippenko}, A.~V., \& {KAIT GRB Team}. 2022, GRB Coordinates
  Network, 32051, 1

\bibitem[{{Zhu} {et~al.}(2022){Zhu}, {Xu}, {Fu}, {Malesani}, {de Ugarte
  Postigo}, {Levan}, \& {Aron}}]{2022GCN.31864....1Z}
{Zhu}, Z., {Xu}, D., {Fu}, S.~Y., {et~al.} 2022, GRB Coordinates Network,
  31864, 1

\end{thebibliography}

\appendix

\section{New telescopes to the GRANDMA collaboration}
\label{tel}

\textcolor{black}{The GRANDMA consortium is a world-wide network telescopes and groups from 18 countries. These facilities make available large amounts of observing time that can be allocated for photometric and/or spectroscopic follow-up of transients located on three continents.}
New telescopes have joined the network since the 2021, during ready-for-O4 campaign as described in \cite{GRANDMA_ztf_fink}. 

Below we describe the new astronomical teams (by alphabetic order) that have contributed to this work.

\paragraph*{C2PU : C2PU-Omicron} -- C2PU is an astronomical research facility located at the Calern site of {\sl Observatoire de la Côte d'Azur} (OCA),
in south-eastern France. The coordinates are $43^\circ\,45^\prime\,13.2^{\prime\prime}$~N, $6^\circ\,55^\prime\,22.7^{\prime\prime}$~E, $1270$~m (above MSL). The median value of the seeing at Calern
is $1.09^{\prime\prime}$ \makeatletter @ \makeatother $\lambda=500$~nm (GDIMM measurements over $3.5$ years; see~\citealt{AristidiEtAl2019a}). The sky background is less than ideal since the site is located within $30$~km from large cities (Cannes, Antibes, Nice). Sky Quality Meter (SQM) background measurements yield values around Vmag=$21$ per squared arc-second.
The C2PU facility is geared with two $1.04$~meter telescopes on equatorial yoke mounts, nicknamed ``Epsilon'' and ``Omicron''. Only Omicron has a wide field mode suitable to participate to the GRANDMA follow-up campaigns. This wide field mode directly uses the focus of the parabolic primary mirror, through a three-lens Wynne coma corrector. The resulting aperture ratio is $F/3.2$. With a QHY600 camera, the FoV is $37.6^\prime\times25.2^\prime$ and the plate scale is $0.47^{\prime\prime}$/pix in $2\times2$ binning mode, and $0.24^{\prime\prime}$/pix in $1\times1$ binning mode.
SDSS filters $g^\prime$, $r^\prime$, and $i^\prime$ are available for this configuration.
The Omicron telescope is in the wide field mode for approximately $40\%$ of time, running GAIA photometric alerts follow-up observations or LSB (Low Surface Brightness) imaging experiments. GRANDMA alerts can be considered as ToO and can be {\textcolor{black}{inserted} within these observation programs.
Data is analyzed by a custom pipeline using GAIA-DR3 catalog for astrometric calibration and SDSS or Pan-STARRS 
catalogs for photometric calibrations.

\paragraph*{GMG} -- The 2.4-meter telescope is located in Gao-Mei-Gu (GMG) village, Lijiang, Yunnan Province, China. The camera called YFOSC has a set of UBVRI (Johnson) filter system and a set of ugriz (sloan) filter system, and one can perform photometric observation. A spectrograph with middle/low spectral resolution is also equipped in the camera such that one can perform spectral observation. The telescope can execute Target-of-Opportunity (ToO) observations. When a trigger is received, a duty member can immediately respond to the trigger where checking target coordinates and telescope conditions is quickly performed. Then, the ToO observation can begin. After the ToO observation, the duty member can access the data catalog and perform preliminary data reduction. GRB observation has been carried out by the telescope since 2010 (as an example, see \citealt{mao12}). The telescope members have joined large international collaborations for GRB observations, GW electromagnetic counterpart and neutrino counterpart searching, and other transient observations. 

\paragraph*{OPD} -- The National  Laboratory of Astrophysics  (LNA) is a research unit of the Brazilian Ministry of Science, Technology and Innovations (MCTI) - that manages the telescopes of the  Pico dos Dias Observatory (OPD).  Located in Braz\'opolis-MG, OPD has provided two telescopes for the GRANDMA campaign collaboration:  a) The Perkin-Elmer 1.6m Telescope (code OPD-1.60m), with  Richey-Chr\'etien optical design and can be used for photometry, polarimetry and spectroscopy equipped with $UBVR_CI_C$ filters; b) The Boller \& Chivens 0.6 m Telescope (code OPD-60) used for photometry and polarimetry, equipped with $UBVR_CI_C$ and $J,H$ filters.

\paragraph*{KAO} -- The Kottamia Astronomical Observatory (KAO) is operated by the National Institute of Astronomy and Geophysics (NRAIG) in Egypt. It is located at 29.9341° N, 31.8277° E, 476 meters (amsl) and is used for scientific observations (\citealt{azzam2010}). KAO telescope has a 1.88-meter primary mirror, equipped with 2kx2k CCD camera covering 8.3 x 8.3 arcmin and with two sets of filters, the SDSS- u'g'r'i'z' and Johnson Cousins-UBVRI (\citealt{Azzam2020}). The KAO team is from NRIAG and accesses the telescope via time proposals. 
For ToO observations, our team is allowed to request the telescope duty on-site observer to observe a GRB counterpart that has a well-defined \textit{Swift} position once the trigger is received.
Time is allocated for the detected GRB counterpart until it fades.
Data reduction and photometry are performed by IRAF and/or Astropy packages using the comparison stars obtained from USNO-B1.0, SDSS, and Pan-STARRS catalogs.

\paragraph*{MOSS, OWL and HAO} -- The Oukaimeden observatory \citep{Benkhaldoun2005} is located 75 km from the city of Marrakesh. The site, located at an altitude of 2750 m, is part of the High Atlas mountain range. It houses several telescopes and other observational instruments \citep{Benkhaldoun2018}. Among these telescopes, three took part in the observation campaigns of the GRANDMA project:
\begin{itemize}
\item MOSS (Moroccan Oukaimeden Sky Survey), is operated within the framework of a cooperation between France, Switzerland and Morocco. It consists of a Newton type telescope (0.5m $F/3$, FOV 1d 23min x 55 min, 3 sigma limit mag 20.8, max 21.5).
\item OWL (Optical Wide-field patroL Network), is operated within the framework of a cooperation between the Korean Astronomy and Space Institute (KASI), and Morocco. The telescope aperture size of the mirror is 0.5m with Ritchey-Chretien configuration, and its field of view is 1.1 deg x 1.1 deg on the CCD sensor. The optical tube assembly was manufactured by Officina Stellare. The alt-az type telescope mount was developed for the OWL-Net exclusively. Its maximum slewing speed is 20 deg/sec, and its acceleration performance is 20 deg/sec. The pointing accuracy is $5 arcsec$, and the tracking accuracy is $2 arcsec/10 minutes$. Although on-tracking is also available, it is just for the experimental use. The system is also equipped with a filter wheel that holds Johnson B, V, R, and I filters.
\item HAO (High Atlas Observatory) is operated within the framework of a cooperation with an association of Moroccan astrophotographers. The telescope used for GRANDMA observations is of the Ritchy Chretien type (CDK 0.318 m $F/8$, FOV 26 sec x 18 sec, 3 sigma limit mag 20.4 in R filter).
\end{itemize}
We incorporate GRANDMA program targets into the operations of the Oukaimeden Observatory based on instrument and observer availability, ensuring no conflicts with previously scheduled programs. This decision is reached in collaboration with our partner teams to maintain program coordination.

\paragraph*{SOAR} -- SOAR is located at 2,700\,m, in Cerro Pach\'on, Chile. It has both optical and near-infrared instruments with field-of-view covering a few arcminutes. Observations are seeing-limited (typical seeing about 0\farcs8). The observations in May were performed using the TripleSpec instrument, a spectrometer that covers the entire 0.9-2.5\, micron region simultaneously. TripleSpec also has a 2\arcmin$\times$2\arcmin Slit View detector operating at the $J$ band, which can also provide scientific data. For GRANDMA, the team also collected broad-band $K$ filter images taken with the Spartan Near-IR Camera. 

\paragraph*{UBAI-AZT-22} -- The AZT-22 telescope of Maidanak Observatory can be used for conducting observations in GRANDMA under the responsibility of the UBAI team. AZT-22 is a 1.5-meter Ritchey–Chrétien telescope and was installed in the late 1980s. The commissioning of the telescope was carried out during 1990–1994 (\cite{Sergeev2014}). The primary mirror, in combination with interchangeable secondary convex hyperbolic mirrors, gives two systems with different image scales with an aperture ratio of 1:7.7 (short focus) and with an aperture ratio of 1:17 (long focus). The main working optical system of AZT-22 is a “short focus” with a length of 11550 mm.
The CCD camera SNUCAM (Seoul National University CAMera) was provided by Seoul National University and installed on AZT-22 in August 2006. It is a camera system with an active array of 4096 by 4096 with physical pixel size of 15 micron (0.268 arcsec). SNUCAM CCD uses UBVRI Bessell filter set \citep{Im2010SEOULNU}. FOV of the telescope with SNUCAM CCD is 18.1'x18.1'. The camera has been cryogenically cooled with CryoTiger closed-cycle refrigeration system. The temperature of the camera was set to -108°C as recommended by the spectral instruments.

\section{Observations from the GRANDMA program}
\label{obsApp}

\paragraph*{Specific time allocation for this campaign} -- During the GRB campaign, special agreements were made with partners to allow possibilities for observation. First, the observation time of AbAO telescopes is divided between three groups of AbAO observers. In the case of ToO observations (GW sources, Lacertids, GRBs, comets, asteroids, etc.), as a rule, we give the telescope to the appropriate group. Since part of the Georgian GRANDMA team has been observing GRBs since 2012 in collaboration with A.S. Pozanenko Group (Space Research Institute RAS), a special agreement was made to obtain the data for this campaign.

The allocation time for the TRT network for the GRB campain was activated using the Target of opportunity time for our submission proposal which is called every 3 months.

In FRAM collaboration, there was a long-lasting GRB observations program outside GRANDMA that is activated automatically upon receiving a trigger over the GCN network and governs the observations of prompt and early afterglow phases of GRBs for up to 2 hours after the trigger.  No events have been observed under this program during Spring 2022 campaign due to unfavorable weather conditions at the sites, or sky positions of the triggers. However, later time observations of GRB220514A have been triggered manually. 

Since being installed, GRB follow-up has been the primary program of VIRT \citep{2004AN....325..669N}. As such, this program is carried out independently of GRANDMA \citep{2021AAS...23713501G}. Over the course of the GRANDMA GRB campaign, though, EO informed GRANDMA of VIRT responses to GRB triggers, and infrastructure was added to monitor the GRANDMA-based alerts systems. 

We obtained some time allocation for observing with WIRCAM mounted on CFHT in 2022A under the name of GRANDMA for 3h time. This time was granted to observe short gamma-ray bursts that may be associated with kilonova and compact binary coalescences.

The data acquired for the afterglow detections under these observations and part of the GRANDMA campaign are reported here.

\begin{table*}
\centering
\caption{Summary of the GRANDMA detected afterglow observations. $\delta \mathrm{t}$ is the delay between the beginning of the observation and the public detection discovery. In this table, only detection magnitudes of each observational epoch are reported. Magnitudes are given in the AB system (calibrated using PS1 or USNOB1) and are not correct for Galactic extinction with calibration.}
\label{tab:KN_observations}
\begin{tabular}{ccccccccccl}
\hline
Source &  Obs date & Time & $\delta \mathrm{t}$(h) & Exposure & Filter & Magnitude & Telescope/ \\
& & & & & &  & Observer \\
\hline
GRB220403B   & 2022-04-03  & 20:50:23 & 0.13 & 5 x 20 s & Clear  & $18.61\pm0.06$  & BJP/ALi-50  \\
GRB220403B   & 2022-04-03  & 20:52:16 & 0.17 & 5 x 20 s & Clear  & $18.73\pm0.07$  & BJP/ALi-50  \\
GRB220403B   & 2022-04-03  & 20:54:10 & 0.2 & 5 x 20 s & Clear  & $18.67\pm0.06$  & BJP/ALi-50  \\
GRB220403B   & 2022-04-03  & 20:56:03 & 0.23 & 5 x 20 s & Clear  & $18.87\pm0.08$ & BJP/ALi-50  \\
GRB220403B   & 2022-04-03  & 21:03:36 & 0.35 & 20 x 20 s & $g'$  & $18.98\pm0.08$ & BJP/ALi-50  \\
GRB220403B   & 2022-04-03  & 21:11:29 & 0.48 & 20 x 20 s & $r'$  & $19.40\pm0.10$  & BJP/ALi-50  \\
GRB220403B   & 2022-04-03  & 21:16:52 & 0.57 & 5 x 20 s & Clear  & $18.84\pm0.08$ & BJP/ALi-50  \\
GRB220403B   & 2022-04-03  & 21:18:44 & 0.6 & 5 x 20 s & Clear  & $19.11\pm0.10$  & BJP/ALi-50  \\
GRB220403B   & 2022-04-03  & 21:20:37 & 0.63 & 5 x 20 s & Clear  & $18.86\pm0.08$ & BJP/ALi-50  \\
GRB220403B   & 2022-04-03  & 21:22:31 & 0.67 & 5 x 20 s & Clear  & $19.09\pm0.09$ & BJP/ALi-50  \\
GRB220403B   & 2022-04-03  & 21:27:38 & 0.75 & 20 x 20 s & $g'$  & $18.95\pm0.08$  & BJP/ALi-50  \\
GRB220403B   & 2022-04-03  & 21:35:29 & 0.88 & 20 x 20 s & $r'$  & $19.94\pm0.15$   & BJP/ALi-50  \\
GRB220403B   & 2022-04-03  & 21:43:42 & 1.02 & 20 x 20 s & Clear  & $19.52\pm0.09$ & BJP/ALi-50  \\
GRB220403B   & 2022-04-03  & 21:51:40 & 1.15 & 20 x 20 s & $g'$  & $19.56\pm0.15$  & BJP/ALi-50  \\
GRB220403B   & 2022-04-03  & 21:59:33 & 1.28 & 20 x 20 s & $r'$  & > 19.6  &  BJP/ALi-50  \\
GRB220403B   & 2022-04-03  & 22:07:47 & 1.42 & 20 x 20 s & Clear  & $19.68\pm0.10$ & BJP/ALi-50  \\
GRB220403B   & 2022-04-03  & 22:11:35 & 1.48 & 20 x 20 s & $r'$  &  $20.11\pm0.12$  & BJP/ALi-50  \\
GRB220403B   & 2022-04-03  & 22:15:44 & 1.55 & 20 x 20 s & $g'$  & > 19.6 &  BJP/ALi-50  \\
GRB220403B   & 2022-04-03  & 22:23:38 & 1.68 & 20 x 20 s & $r'$  & > 19.8   &  BJP/ALi-50  \\
GRB220403B   & 2022-04-03  & 22:31:51 & 1.81 & 20 x 20 s & Clear  & $19.91\pm0.11$  & BJP/ALi-50  \\

GRB220403B   & 2022-04-04  & 13:44:06 & 17.67 & 15 x 300 s & Clear  & > 21  &  SNOVA  \\

GRB220403B   & 2022-04-04  & 20:40:18 & 23.97 & 300 s & $R$  & $22.22\pm0.12$   &  UBAI/AZT-22  \\

GRB220403B   & 2022-04-04  & 21:25:43 & 24.72 & 24 x 32 s & $B$ & > 21.0  &  KNC-T-CAT \\
GRB220403B   & 2022-04-04  & 21:25:43 & 24.72 & 24 x 32 s & $G$  & > 21.1  &  KNC-T-CAT  \\
GRB220403B   & 2022-04-04  & 21:25:43 & 24.72 & 24 x 32 s & $R$  & > 20.9  &  KNC-T-CAT  \\

GRB\,220427A & 2022-04-27 & 21:04:04 & 0.05833 & 30 s & Clear & $16.32\pm0.22$ & TAROT/TRE \\
GRB\,220427A & 2022-04-27 & 21:04:43 & 0.06916 & 30 s & Clear & $15.74\pm0.18$ & TAROT/TRE \\
GRB\,220427A & 2022-04-27 & 21:05:21 & 0.07972 & 30 s & Clear & $16.04\pm0.18$ & TAROT/TRE \\
GRB\,220427A & 2022-04-27 & 21:05:59 & 0.09028 & 30 s & Clear & $16.33\pm0.19$ & TAROT/TRE \\
GRB\,220427A & 2022-04-27 & 21:06:37 & 0.10083 & 30 s & Clear & $16.51\pm0.16$ & TAROT/TRE \\
GRB\,220427A & 2022-04-27 & 21:07:36 & 0.11722 & 30 s & Clear & $16.77\pm0.11$ & TAROT/TRE \\
GRB\,220427A & 2022-04-27 & 21:09:15 & 0.14472 & 30 s & Clear & $17.26\pm0.35$ & TAROT/TRE \\
GRB\,220427A & 2022-04-27 & 21:17:26 & 0.28111 & 30 s & Clear & $18.0\pm0.3$ & TAROT/TRE \\
GRB\,220427A & 2022-04-27 & 22:08:04 & 1.1250 & $15\times120$ s & Clear & $20.09\pm0.12$ & Les Makes/T60 \\
GRB\,220427A & 2022-04-27 & 22:38:32 & 1.6327 & $15\times120$ s & Clear & $20.44\pm0.14$ & Les Makes/T60 \\
GRB\,220427A & 2022-04-27 & 23:41:31 & 2.6825 & $15\times120$ s & Clear & $21.69\pm0.22$ & Les Makes/T60 \\
GRB\,220427A & 2022-04-27 & 00:52:03 & 3.8581 & $15\times120$ s & Clear & $21.64\pm0.21$ & Les Makes/T60 \\

GRB220514A   & 2022-05-14  & 13:58:35 & 1.57 & 3 x 300 s & $r^\prime$  & $18.9\pm0.1$   & Xinglong-TNT  \\

GRB220514A   & 2022-05-14  & 20:04:11 & 7.66 & 9 x 60 s & $R$  &  > 17.3  & Abastumani/T70 \\

GRB220514A   & 2022-05-14  & 20:21:43 & 7.95 & 45 x 60 s & Clear &  $19.9 \pm 0.25$ & MOSS \\

GRB220514A   & 2022-05-14  & 20:38:27 & 8.23 & 15 x 120 s & $R$ & >19.7  & KNC-HAO \\

GRB220514A   & 2022-05-14  & 20:40:25 & 8.50 & 15 x 120 s & Clear  & > 19  & FRAM-CTA-N \\

GRB220514A   & 2022-05-14  & 21:59:16 & 9.58 & 10 x 150 s & $i'$  & $20.3 \pm 0.1 $ & 2.2m CAHA/CAFOS \\

GRB220514A   & 2022-05-15  & 03:27:54 & 15.06 & 17 x 180 s & $Rc$ & > 19.6 & KNC-iT21 \\

\end{tabular}
\end{table*}

\begin{table*}
\centering
\caption{Selection of one observation per telescope that images were provided during the campaign and allowed us to provide an upper limit. Here we present 17 telescopes that provided good-quality images resulting in upper limits.}
\label{tab:selected_observations}
\begin{tabular}{ccccc}
\hline
Telescope &  Source & $\delta \mathrm{t}$(h) & Limiting magnitude & Filter\\
\hline
BJP/ALi-50   & GRB220403B  & 0.14 & 20.1 & Clear\\
SNOVA   & GRB220430A  & 17.74 & 19.3 & Clear\\
NOWT   & GRB220325A  & 4.72 & 20.9 & $R$\\
KAO   & GRB220325A  & 8.11 & 21.5 & $R$\\
C2PU-Omicron   & GRB220325A  & 10.16 & 21.6 & $r'$\\
UBAI/AZT-22   & GRB220412A  & 33.9 & 21.7 & $R$\\
TRT-SRO   & GRB220408A  & 1.82 & 19.3 & $R$\\
Xinglong-TNT   & GRB220408A  & 4.95 & 20.6 & $R$\\
MOSS & GRB220430A  & 6.59 & 20.3 & Clear\\
GMG-2.4   & GRB220412A  & 6.06 & 20.5 & $r'$\\
VIRT   & GRB220412A  & 17.59 & 20.3 & $R$\\
Les Makes/T60  & GRB220427A  & 1.13 & 20.4 & Clear\\
TAROT/TRE  & GRB220427A  & 0.06 & 16.4 & Clear\\
Abatsumani/T70  & GRB220430A  & 3.22 & 18.7 & $R$\\
OWL   & GRB220514A  & 1.82 & 18.4 & Clear\\
FRAM-CTA-N   & GRB220514A  & 2.28 & 18.3 & Clear\\
2.2mCAHA   & GRB220514A  & 9.61 & 22.0 & $i'$\\
\end{tabular}
\end{table*}

\end{document}